\documentclass[usenatbib,useAMS,usedcolumn]{mn2e}
\usepackage{times,epsfig}
\pdfoutput=1

\newcommand{\arcm}{\hbox{$^\prime$}}

\newcommand{\degree}{\hbox{$^\circ$}}
\newcommand{\rosat}{\emph{ROSAT}}
\newcommand{\chandra}{\emph{Chandra}}
\newcommand{\xmm}{\emph{XMM--Newton}}
\newcommand{\xmms}{\emph{XMM}}
\newcommand{\asca}{\emph{ASCA}}

\newcommand{\arcs}{\mbox{\arcm\arcm}}
\newcommand{\Lx}{\ensuremath{L_{\mathrm{X}}}}

\newcommand{\Zsol}{\ensuremath{~Z_{\odot}}}

\newcommand{\Msol}{\ensuremath{~M_{\odot}}}
\newcommand{\LB}{\ensuremath{L_{\mathrm{B}}}}
\newcommand{\LBsol}{\ensuremath{L_{B\odot}}}
\newcommand{\LKsol}{\ensuremath{L_{K\odot}}}

\newcommand{\s}{\ensuremath{\mbox{~s}}}
\newcommand{\ps}{\ensuremath{\s^{-1}}}
\newcommand{\cm}{\ensuremath{\mbox{~cm}}}
\newcommand{\pcmsq}{\ensuremath{\cm^{-2}}}
\newcommand{\pcmcu}{\ensuremath{\cm^{-3}}}
\newcommand{\km}{\ensuremath{\mbox{~km}}}
\newcommand{\Mpc}{\ensuremath{\mbox{~Mpc}}}
\newcommand{\pMpc}{\ensuremath{\Mpc^{-1}}}
\newcommand{\kmpspMpc}{\ensuremath{\km \ps \pMpc\,}}
\newcommand{\erg}{\ensuremath{\mbox{~erg}}}
\newcommand{\ergps}{\ensuremath{\erg \ps}}
\newcommand{\ergpspcmsq}{\ensuremath{\erg \ps \pcmsq}}
\newcommand{\ergpcmcu}{\ensuremath{\erg \pcmcu}}
\newcommand{\kmps}{\ensuremath{\km \ps}}

\newcommand{\sas}{\textsc{sas}}

\newcommand{\Ho}{\ensuremath{\mathrm{H_0}}}

\newcommand{\Dtf}{\ensuremath{D_{\mathrm{25}}}}

\begin{document}

\title[ Intergalactic medium / radio source interaction in NGC~4261 ]{Interaction between the intergalactic medium and central radio source in the NGC~4261 group of galaxies}

\author[E. O'Sullivan et al.]
{E. O'Sullivan$^{1,2}$\thanks{ejos@star.sr.bham.ac.uk},
  D.~M. Worrall$^{3}$, M. Birkinshaw$^{3}$, G. Trinchieri$^{4}$,
  A. Wolter$^{4}$, \newauthor A. Zezas$^{2,5,6}$ and S. Giacintucci$^{2,7}$\\
$^1$ School of Physics and Astronomy, University of Birmingham, Birmingham, B15 2TT, UK \\
$^2$ Harvard-Smithsonian Center for Astrophysics, 60 Garden
  Street, Cambridge, MA 02138, USA \\
$^3$ HH Wills Physics Laboratory, University of Bristol, Tyndall Avenue, Bristol BS8 1TL, UK\\
$^4$ Osservatorio Astronomico di Brera, via Brera 28, 20121 Milano, Italy \\
$^5$ Physics Department, University of Crete, Heraklion, GR-71003, Greece \\
$^6$ IESL, Foundation for Research and Technology - Hela, 7110 Heraklion,
Greece\\
$^7$ Department of Astronomy, University of Maryland, College Park, MD
20742-2421, USA\\
}

\date{Accepted 2011 June 13; Received 2011 June 02; in original form 2011 March 17}

\pagerange{\pageref{firstpage}--\pageref{lastpage}} \pubyear{2010}

\label{firstpage}

\maketitle

\begin{abstract}
  Using observations from the \chandra\ and \xmm\ X-ray observatories, we
  examine the interaction between the intra--group medium and central radio
  source in the nearby NGC~4261 galaxy group. We confirm the presence of
  cavities associated with the radio lobes and estimate their enthalpy to
  be $\sim$2.4$\times10^{58}$~erg. The mechanical power output of the jets
  is $\geq$10$^{43}$\ergps, at least a factor of 60 greater than the
  cooling luminosity in the region the lobes inhabit. We identify rims of
  compressed gas enclosing the lobes, but find no statistically significant
  temperature difference between them and their surroundings, suggesting
  that the lobe expansion velocity is approximately sonic
  ($\mathcal{M}\leq1.05$). The apparent pressure of the radio lobes, based
  on the synchrotron minimum energy density argument, is a factor of 5
  lower than that of the intra--group medium.  Pressure balance could be
  achieved if entrainment of thermal gas provided additional non-radiating
  particles in the lobe plasma, but the energy required to heat these
  particles would be $\sim$20 per cent. of the mechanical energy output of
  the radio source. NGC~4261 has a relatively compact cool core, which
  should probably be categorised as a galactic corona. The corona is
  capable of fuelling the active nucleus for considerably longer than the
  inferred source lifetime, but can be only inefficiently heated by the AGN
  or conduction. The expansion of the radio lobes has affected the
  structure of the gas in the galaxy, compressing and moving the material
  of the corona without causing significant shock heating, and expelling
  gas from the immediate neighbourhood of the jets.  We discuss the
  possible implications of this environment for the duration of the AGN
  outburst, and consider mechanisms which might lead to the cessation of
  nuclear activity.
\end{abstract}


\section{Introduction}
\label{sec:intro}
The importance of radio galaxies as sources of heating in groups and
clusters of galaxies has become increasingly apparent in recent years.
Active galactic nuclei (AGN) are now considered the most likely mechanism
acting to balance radiative cooling of the hot intra--group medium
\citep[IGM,][]{McNamaraNulsen07,PetersonFabian06}. Approximately 50 per
cent. of galaxy clusters \citep[e.g.,][]{Sandersonetal06} and perhaps as
many as 85 per cent. of galaxy groups \citep{Dongetal10} have a cooling
region centred on a giant elliptical or cD galaxy, and the link between
cool cores and nuclear activity in these galaxies is well established
\citep[e.g.,][]{Burns90,Mittaletal09,Sun09}.

X--ray and radio imaging has provided a means to examine the interaction
between radio galaxies and their environment, and revealed that many nearby
groups and clusters contain complex structures associated with the radio
jets and lobes. Shocks and sound waves driven by the expansion of radio
jets may directly heat the gas
\citep[e.g.,][]{Nulsenetal05,Fabianetal06,SandersFabian07,Formanetal07}
while cavities may inject energy by doing work on the gas, or reduce its
ability to cool by lifting material out from the densest, most highly
enriched regions \citep[e.g.,][]{Fabianetal05,Wiseetal07,Blantonetal09}.
The enthalpy of cavities associated with radio lobes has been shown to be
sufficient to balance radiative cooling in many systems
\citep{Birzanetal08}, and there is evidence that in some systems shocks may
deposit similar amounts of energy into the gas
\citep[e.g.,][]{Millionetal10}.

However, the mechanisms by which AGN are fuelled, and thus the processes
involved in starting and stopping nuclear activity are not clear.
Correlations between the AGN power output and the Bondi accretion rate
suggest that Fanaroff-Riley class I (FR\,I) radio sources could be fuelled
by hot gas \citep{Allenetal06,Hardcastleetal07,Balmaverdeetal08}. While
actual Bondi accretion would require unrealistic efficiencies
\citep{McNamaraetal11}, this at least suggests that fuelling by gas cooled
from the hot phase is a possibility. Fuelling by large reservoirs of cold
molecular gas or neutral hydrogen appears to be less likely, with poor
correlation between cold gas mass and radio power
\citep{McNamaraetal11,Emontsetal07}. To form a feedback system, the AGN
must be able to effectively heat its fuel supply, located in its immediate
surroundings. Shocks originating in the nuclear region could be effective
in this regard, while cavity heating would require conduction to distribute
the energy azimuthally and perhaps radially.

While these mechanisms may be feasible in many systems, they appear
difficult to achieve in those dominant ellipticals which host galactic
coronae, cool cores with radii of only a few kiloparsecs. Coronae are
thought to be largely unaffected by the radio jets of the AGN they host,
and conduction into the cool gas from the surrounding IGM is suppressed
\citep{Sunetal07}. Owing to their small size, relatively small heating
efficiencies ($<$1 per cent. in some cases) would raise the temperature of
the corona gas to that of the ambient medium, suggesting that conduction is
strongly suppressed by magnetic fields, and that the jets tunnel through
the cool core with little or no interaction.  Coronae are thought to
consist of gas lost from stars within the cool region. Their cooling rate
is often similar to the rate of stellar mass loss in the core
\citep{Sunetal07}, implying that if the AGN is fuelled by cooling gas, a
corona can fuel its AGN over long periods without requiring inflowing gas
from the IGM. Destruction of coronae by mergers or by ram--pressure
stripping seems infeasible given the relatively large numbers of such
systems observed \citep{Sunetal07}. Coronae thus raise serious problems for
the feedback model, since they can apparently provide enough gas to fuel an
AGN for long periods, without being heated by the jets this activity
produces.  While this possibility is as yet speculative, there is at least
one example of a corona fuelling an unusually old radio galaxy
\citep{OSullivanetal10}. AGN in corona systems are also known to have
different radio properties; the radio power of radio galaxies in large cool
cores scales with the luminosity of the cool core, whereas those in coronae
show no such relationship \citep{Sun09}.

The question of how the supply of gas to the AGN can be stopped is
important in all cluster and group--central AGN, and particularly so in
coronae. Addressing this issue is difficult in that it requires examination
of the AGN/gas interaction on very small scales. Many cavity systems
studied to date are distant, making such an examination difficult, and this
is particularly true of corona systems. It would therefore be useful to
study nearby radio galaxies embedded in small-scale cool cores or coronae,
to determine what effect the radio source has on the core, and how this
might affect fuelling of the AGN.

One such system is the NGC~4261 group, which has a well-established cool
core and whose central elliptical hosts the well known FR\,I radio source
3C~270. Our aim in this paper is to use the available deep \chandra\ and
\xmm\ observations of the group to examine the interaction between the
radio source, the cool core and the surrounding intra-group medium. We can
thus determine the effects of the ongoing AGN outburst (or series of
outbursts) on the gas, the energies and timescales involved, and the
current physical structure of the gas.  NGC~4261 is particularly suited to
such a study, since the well--defined axis of its jets removes any question
of large projection effects in determining the size of structures, and its
radio properties and black hole mass are relatively well constrained. We
describe the group, galaxy and radio source in more details in
Section~\ref{sec:4261}. In Section~\ref{sec:obs} we describe the \chandra\
and \xmm\ observations and their reduction, and in Section~\ref{sec:res}
the results of imaging and spectral analysis of these data, including
estimates of the age and expansion rate of the cavities. We further analyse
these results in Section~\ref{sec:ana} to examine the energy output of the
AGN, the properties and particle content of the radio lobes, and the state
of the cool core. These results and their relationship with, and
implications for, other group-- and cluster--central radio galaxies are
discussed in Section~\ref{sec:discuss}, and we summarise our conclusions in
Section~\ref{sec:conc}.

Throughout this paper we assume \Ho=70\kmpspMpc\ and a distance for
NGC~4261 of 31.6~Mpc, in line with that adopted by \citet{Worralletal10}.
This gives an angular scale of 1\arcm\ = 9.2~kpc. We note that
\citet{Humphreyetal09b} use a smaller distance (29.3~Mpc) based on surface
brightness fluctuation measurements \citep{Tonryetal01}. Our conclusions
would not be altered were we to adopt this distance value.
\citet{Worralletal10} also adopt the redshift of NGC~4261, 0.00746
\citep{Trageretal00}, as appropriate for their goal of studying the AGN and
jets. As our aims concern the IGM, we adopt the mean redshift of the group,
0.00706 \citep{Nolthenius93}. However, we note that adopting the galaxy
redshift would not significantly alter the fit results.

\subsection{NGC 4261}
\label{sec:4261}
NGC~4261 and its surrounding group have been extensively studied, and it is
useful to review what is known of the system before moving on to discuss
our analysis. NGC~4261 is a boxy, cuspy--cored , slowly rotating E2 galaxy
\citep{Ravindranathetal01}, whose primary axis of rotation is close to the
major axis, leading to the conclusion that the galaxy is prolate
\citep{DaviesBirkinshaw86}. The galaxy hosts a low-ionisation nuclear
emission-line region \citep[LINER,][]{Hoetal97b} and the bright FR\,I radio
source 3C~270 \citep[$\sim$19~Jy at 1.4~GHz,][]{Kuhretal81}, whose twin
jets lie close to the plane of the sky.  The galaxy has a kiloparsec--scale
kinematically decoupled core \citep{Cappellarietal07} which contains a
100~pc--scale disk of dust and cool molecular and atomic gas
\citep{Jaffeetal93,Jaffeetal94} whose rotation axis is closely matched with
the axis of the radio jets \citep{Ferrareseetal96}. On large scales, deep
optical imaging has revealed a faint tidal tail to the northwest and tidal
fan extending southeast from the galaxy \citep{Taletal09} and there is
evidence of anisotropy in the globular cluster distribution
\citep{Giordanoetal05}. These disturbed features suggest that NGC~4261
underwent tidal interactions or merger with another galaxy within the past
1-2~Gyr.

Diffuse X--ray emission was first detected in the NGC~4261 group using the
\rosat\ PSPC, which revealed a gaseous halo extending to at least 40\arcm\
($>$360~kpc) with a luminosity weighted mean temperature of $\sim$0.85~keV
\citep{Davisetal95,WorrallBirkinshaw94}. \asca\ observations confirmed the
presence of a central, spectrally hard X--ray source
\citep{Matsumotoetal01}. The poor spatial resolution of these data
prevented any detailed examination of structure within the halo, but more
recent observations by \chandra\ and \xmm\ have revealed X--ray features
which are clearly related to the radio source and its interaction with its
environment. On scales of a few arcminutes, \xmm\ imaging showed arm-like
features enclosing the inner edges of the radio lobes
\citep{Crostonetal05}, and confirmed that at least in the western lobe,
these correspond to the edges of a cavity \citep{Crostonetal08}.

On smaller scales, a short \chandra\ exposure revealed X--ray jets
corresponding to the inner few kiloparsecs of the radio jets
\citep{Gliozzietal03,Zezasetal05}, strongly indicating that the jet axis is
close to the plane of the sky. Analysis of a more recent, longer
observation suggests a synchrotron origin for the jet X--ray emission, and
finds wedge-like regions of reduced surface brightness along the jet axis
\citep{Worralletal10}. These are interpreted as conical volumes in which
the thermal gas has been displaced by relativistic plasma from the radio
lobes, suggesting that up to 20 per cent. of the gas within $\sim$10 kpc of
the galaxy core has been moved by the action of the AGN jets and lobes.
Radial spectral analysis confirms that the group temperature profile
follows a typical form, with a central cool core ($kT\sim$0.6~keV) of
radius $\sim$10~kpc, at the upper end of the range observed in other corona
systems, and a temperature peak at $\sim$200\arcs\ ($kT\sim$1.6~keV), with
a decline at larger radii \citep{Helsdonponman00,Humphreyetal06}. These
profiles have been used to place limits on the total mass of the group
\citep[$\sim$6$\times10^{13}$\Msol,][]{Humphreyetal06}, and on the central
supermassive black hole ($\sim$4.4$\times10^8$\Msol) which are in good
agreement with dynamical estimates \citep{Humphreyetal09,Ferrareseetal96}.

\section{Observations and Data Reduction}
\label{sec:obs}
\subsection{Chandra}
NGC~4261 has been observed twice by the \chandra\ ACIS instrument, first
during Cycle~1 on 2001 May 26 (ObsId 894),
for $\sim$35~ks, using 1/2 subarray mode, and again in Cycle~9 on 2008
February 12 (ObsId 9569)
for just over
100~ks.  A summary of the \chandra\ mission and instrumentation can be
found in \citet{Weisskopfetal02}. In both observations the S3 CCD was
placed at the focus of the telescope and the instrument operated in faint
mode for the first observation and very faint mode for the second. We have
reduced the data from the pointings using CIAO 4.1.2 and CALDB 4.1.3
following techniques similar to those described in \citet{OSullivanetal07}
and the \chandra\ analysis
threads\footnote{http://asc.harvard.edu/ciao/threads/index.html}.  The
level 1 events files were reprocessed, very faint mode filtering was
applied to the second dataset, bad pixels and events with \asca\ grades 1,
5 and 7 were removed, and the cosmic ray afterglow correction was applied.
The data were corrected to the appropriate gain map, the standard
time-dependent gain and charge-transfer inefficiency (CTI) corrections were
made, and a background light curve was produced.  The observations did not
suffer from significant background flaring, and the final cleaned exposure
times were 29.6 and 100.9~ks. While data from the entire detector were
examined, for the purposes of this study we primarily use the S3 CCD, as
the radio source and the galaxy fall on that chip.

Identification of point sources on S3 was performed using the \textsc{ciao}
task \textsc{wavdetect}, with a detection threshold of 10$^{-6}$, chosen to
ensure that the task detects $\leq$1 false source in the field, working
from a 0.3-7.0 keV image and exposure map from the combined observations.
Source ellipses were generated with axes of length 4 times the standard
deviation of each source distribution. These were then used to exclude
sources from most spectral fits. An extended source was detected coincident
with the peak of the diffuse X-ray emission; this was not excluded.

Spectra were extracted using the \textsc{specextract} task. Spectral
fitting was performed in XSPEC 12.6.0. Abundances were measured relative to
the abundance ratios of \citet{GrevesseSauval98}. A galactic hydrogen
column of 1.75$\times10^{20}$\pcmsq, drawn from the \textsc[ftools] task nh
and based on the survey of \citet{Kalberlaetal05}, was adopted in all fits.
This differs slightly from the column of 1.58$\times10^{20}$\pcmsq adopted
by \citet{Worralletal10}, based on the survey of \citet{DickeyLockman90}.
The difference probably arises from the finer angular resolution of the
Kalberla et al. survey (0.675$\degree$ compared to 1$\degree$), but testing
suggests it has no significant effect on our spectral fitting. 90 per cent
errors are reported for all fitted values. Spectra were grouped to 20
counts per bin, and counts at energies above 7 keV and below 0.7~keV (see
below) were ignored during fitting.

Background spectra were drawn from the standard set of CTI-corrected ACIS
blank sky background events files in the \chandra\ CALDB. The exposure time
of each background events file was altered to produce the same 9.5-12.0 keV
count rate as that in the target observation. A region enclosing the AGN
and jets was excluded from the estimation of the count rate of the target
observation, to avoid any contamination from source photons.  Very faint
mode background screening was applied to the background data sets where
appropriate.  Comparison of source and background spectra suggested some
mismatch between the source and background spectra, mainly below 0.5 keV.
This is not unexpected, as the soft X-ray background arises largely from
hot gas in our galaxy, and from coronal emission associated with solar wind
interactions, and thus is both spatially and temporally variable
\citep[e.g.,][]{KuntzSnowden00,Snowdenetal04}. There are also indications
that the spectral shape of the background has changed since the creation of
the blank-sky background files (c.f. the ACIS background
cookbook\footnote{http://asc.harvard.edu/contrib/maxim/acisbg/COOKBOOK}),
which could contribute to the disagreement at low energies. NGC~4261 also
lies on the outskirts of the Virgo cluster and close to the galactic north
polar spur \citep{Bohringeretal94}. It is possible that emission from both
these sources could contaminate our observations. We therefore ignored
energies below 0.7 keV when performing spectral fitting, so as to avoid any
biases arising from inaccuracies in estimating the soft X--ray background.

\subsection{XMM-Newton}
\label{sec:xmm}
The NGC~4261 group was observed with \xmm\ during Cycle 1 (2001 December
16) for just over 33~ks (ObsId 0056340101) and again in Cycle~6 (2007
December 16 and 18) for a total of 130~ks (ObsIds 0502120101 and
0502120201). We reduced all three observations, but as ObsId 0502120101 has
by far the longest exposure ($\sim$127~ks before filtering) our analysis
focussed on this dataset.

The EPIC instruments were operated in full frame mode, with the medium
optical blocking filter. A detailed summary of the \xmm\ mission and
instrumentation can be found in \citet[and references
therein]{Jansenetal01}. Reduction and analysis were performed using
techniques similar to those described in \citet{OSullivanetal07}.  The raw
data from the EPIC instruments were processed with the \xmm\ Science
Analysis System (\textsc{sas v.9.0.0}), using the \textsc{epchain} and
\textsc{emchain} tasks. Bad pixels and columns were identified and removed,
and the events lists filtered to include only those events with FLAG = 0
and patterns 0-12 (for the MOS cameras) or 0-4 (for the PN). The total
count rate for the field revealed significant background flaring. Times
when the total count rate deviated from the mean by more than 3$\sigma$
were therefore excluded.  The effective exposure times for the MOS and PN
cameras were 72.9 and 46.8 ksec respectively for ObsId 0502120101.

Images and spectra were extracted from the cleaned events lists with the
\sas\ task \textsc{evselect}. Response files were generated using the \sas\
tasks \textsc{rmfgen} and \textsc{arfgen}. The central AGN is relatively
X-ray bright, and a significant number of out-of-time (OOT) events are
found on the PN detector, visible in the PN data as a trail extending from
the centre of the source toward the CCD readout. An OOT events list was
created using \textsc{epchain}, and scaled by 0.063 to allow statistical
subtraction of the OOT events from spectra and images (c.f. the \xmm\ users
handbook\footnote{http://xmm.esac.esa.int/external/xmm\_user\_support/documentation/}).
Point sources were identified using \textsc{edetect$\_$chain}, and regions
corresponding to the 85 per cent encircled energy radius of each source
were excluded. A source corresponding to the active nucleus was not
excluded.

Creation of background images and spectra for the system was hampered by
the fact that the group X-ray halo extends beyond the field of view. Use of
the ``double-subtraction'' technique \citep{Arnaudetal02,Prattetal01}
involves correcting blank-sky exposures to match a source-free area of the
observation; for NGC~4261 this is not feasible. We therefore adopt several
approaches. For deprojected radial spectral profiles, we use a local
background extracted at the extreme edge of the field of view and discard
results from the outer annulus of the radial profile as potentially biased.
The effect of the deprojection should act to reduce any bias in the
successive inner annuli. Comparison with our deprojected \chandra\ profile
and with that of \citet{Humphreyetal09b} shows good agreement, suggesting
that our results are not seriously affected. For spectral analysis of
particular X-ray structures we use local background spectra extracted close
to the region of interest, so as to subtract off any overlying group and
background emission.

\subsection{Very Large Array}
\label{sec:VLA}
We analysed 1.5 GHz data in C-array configuration retrieved from the Very
Large Array (VLA) public archive (project AL693). The observations were
made in two 25-MHz bands centred on 1365~MHz and 1646~MHz, in May 2008 for
a total integration time on source of approximately 38 minutes. We used the
NRAO Astronomical Image Processing System (AIPS) package for the data
reduction and analysis. Data calibration and imaging were carried out
following the standard procedure (Fourier Transform, Clean and Restore).
Phase-only self-calibration was applied to remove residual phase variations
and improve the quality of the image. The final image has an angular
resolution of 16\arcs$\times$ 15\arcs\ and a rms noise level ($1\sigma$) of
0.3 mJy~beam$^{-1}$.

\section{Results}
\label{sec:res}

\subsection{IGM structures associated with the radio source}
\label{sec:im2}
We initially examined images extracted from both \chandra\ and \xmm,
smoothed at a range of scales, to determine the distribution of diffuse
emission. The AGN is the brightest source in the field, and the jets are
clearly visible in the \chandra\ data. The diffuse emission is strongly
centrally peaked, and fills both the \chandra\ and \xmm\ fields of view. 

\begin{figure*}
\centerline{
\includegraphics[width=\columnwidth, viewport=35 170 577 622]{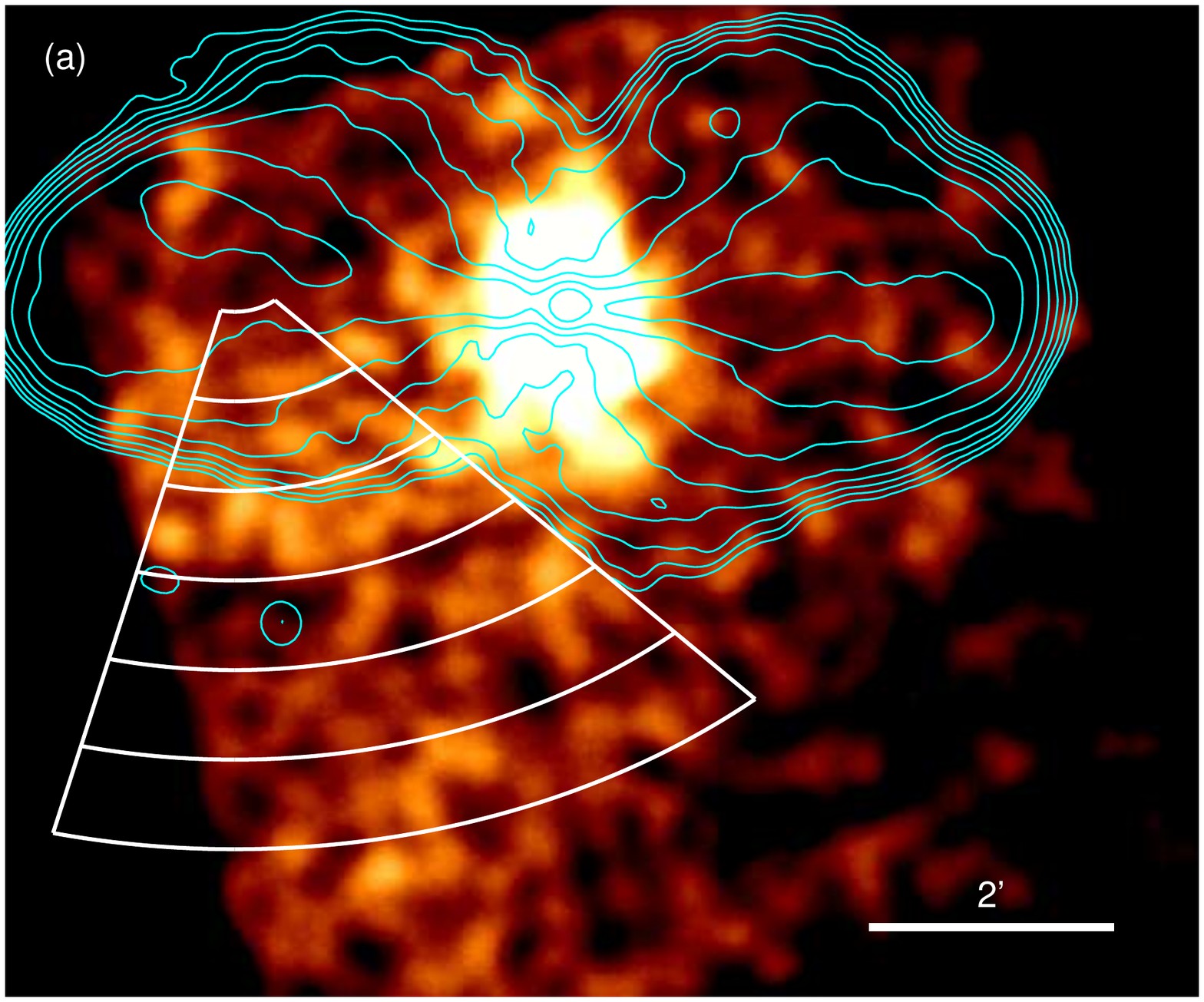}
\includegraphics[width=\columnwidth, viewport=35 170 577 622]{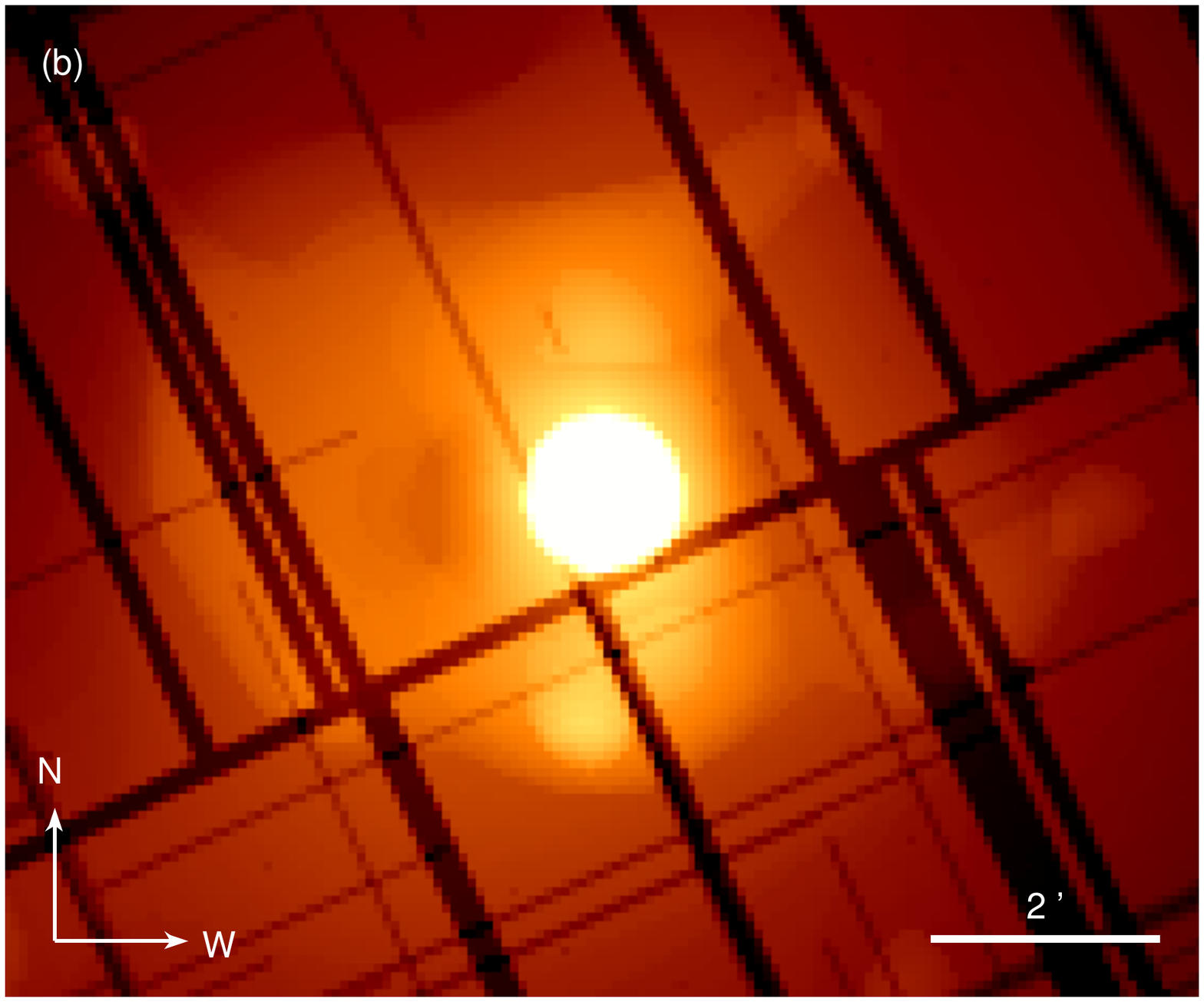}
}
\centerline{
\includegraphics[width=\columnwidth, viewport=35 170 577 622]{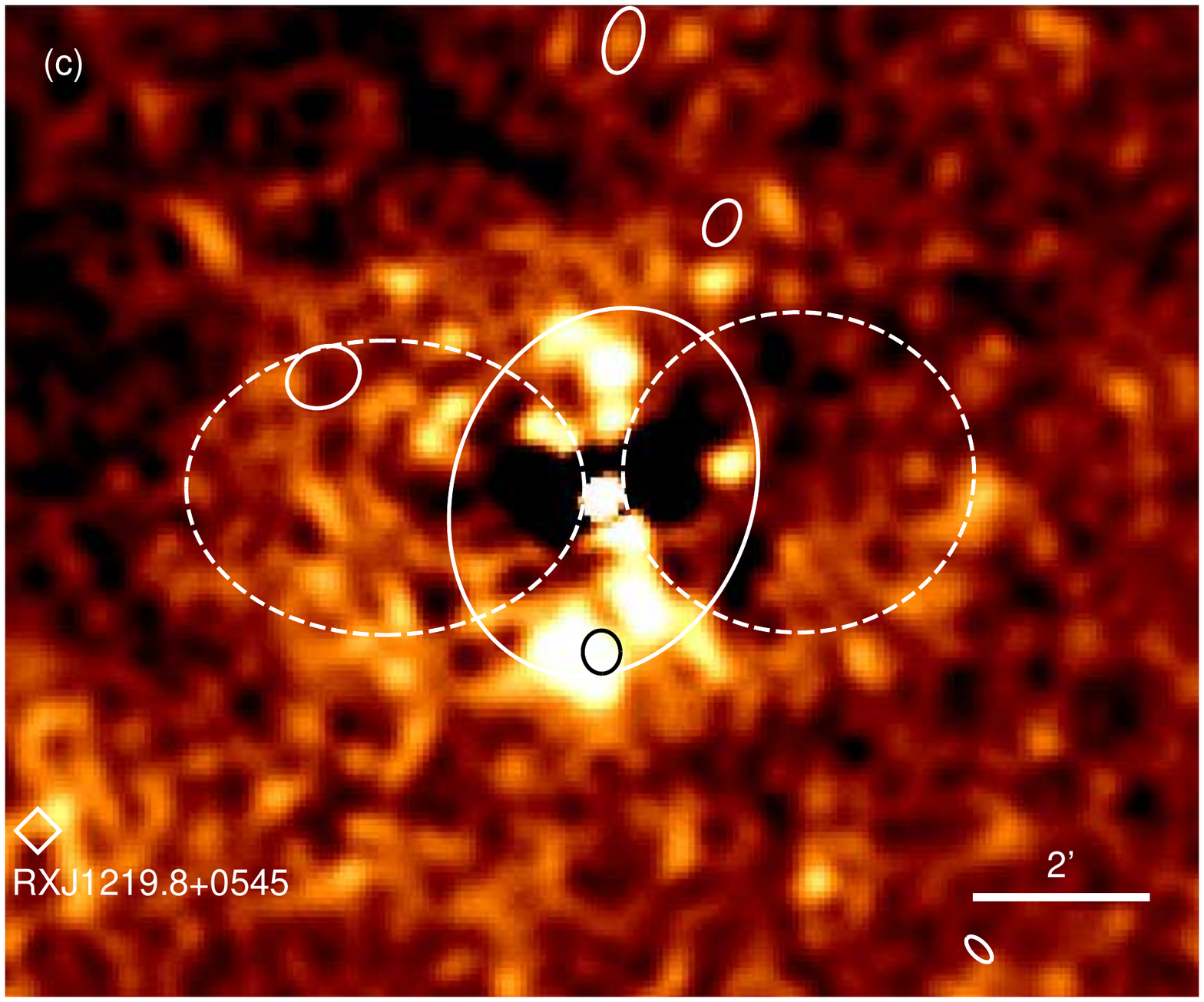}
\includegraphics[width=\columnwidth, viewport=35 170 577 622]{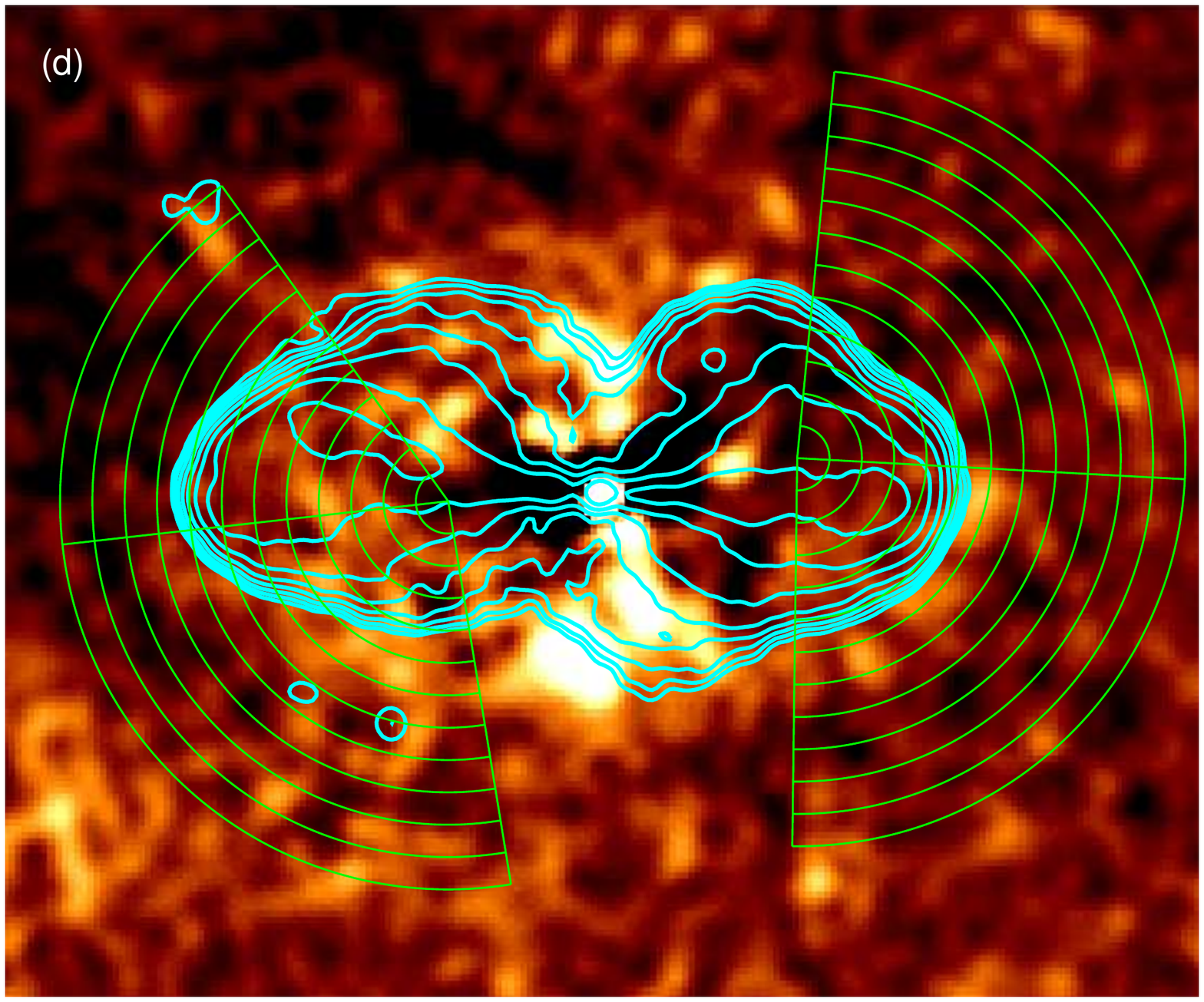}
}
\caption{\label{fig:4ims} \textit{a)} \chandra\ 0.3-3~keV image, binned by
  a factor of 2 and smoothed with a Gaussian of sigma 16 pixels
  (15.8\arcs). VLA 1.5~GHz contours are overlaid, starting 3$\sigma$ above
  the r.m.s. noise level of 0.3 mJy beam$^{-1}$ and increasing by factors
  of 2. The restoring beam size is 16\arcs$\times$15.1\arcs\ (HPBW).
  Annular regions used to examine surface brightness across the cavity rim
  are marked in white. \textit{b)} \xmm\ 0.3-3~keV wavelet smoothed MOS+pn
  image, with scales chosen to exclude point sources and emphasise diffuse
  emission. Dark linear features show chip gaps and bad columns, and should
  be ignored.  \textit{c)} \xmm\ residual map, showing the image after
  removal and filling of point sources, and subtraction of the best fitting
  surface brightness model for the galaxy and larger group halo, smoothed
  with a Gaussian of sigma 4 pixels (17.6\arcs). Solid ellipses indicate
  the optical \Dtf\ contours of NGC~4261 and several smaller galaxies in
  the field.  Dashed ellipses indicate the regions used to define cavity
  size and position. \textit{d)} \xmm\ residual map, as in the lower left
  panel, overlaid with VLA contours and the partial annuli used to extract
  surface brightness profiles across the cavity rims.}
\end{figure*}

On scales of a few arcminutes, similar to the size of the stellar component
of the galaxy (see Figure~\ref{fig:4ims}a), it is clear that moderately
luminous diffuse emission extends north and south of the central AGN, to a
distance of at least 2\arcm\ ($\sim$18~kpc). This extension is
perpendicular to the jet axis, and poorly aligned with the major axis of
the galaxy. On larger scales, fairly uniform emission extends throughout
the field of view of both instruments. However, comparing archival VLA
1.5~GHz radio maps to the X-ray images, there are hints of structures
associated with the radio lobes. In particular, after subtraction of point
sources from the \chandra\ image and refilling of the resulting holes using
the \textsc{ciao} \texttt{dmfilth} task, heavy smoothing reveals an
apparent arc of X--ray emission extending along the southern edge of the
eastern radio lobe, and hints of a similar structure south of the western
lobe (see Figure~\ref{fig:4ims}a).  These could be parts of a shell of
compressed gas surrounding cavities excavated by the expanding radio lobes,
but the limited field of view of the S3 CCD prevents us from determining
whether the structures completely enclose the lobes.

The \xmm\ images are large enough to cover the whole radio source, but are
more severely affected by the bright AGN emission, owing to the broader
\xmm\ point spread function. Other point sources in the galaxy, and sources
associated with other galaxies in the field of view, are also problematic,
and unresolved sources increase the level of noise in the area of interest.
We therefore adopted two approaches aimed at clearly determining whether
there are cavities corresponding to the radio lobes, and the overall form
of any structures in the IGM. As a simple test, we applied the
\texttt{wvdecomp} wavelet decomposition and smoothing algorithm
\citep{Vikhlininetal98} to examine the amount of structure on different
spatial scales. Removal of features with small smoothing scales ($\leq$3
pixels or 13.2\arcs) allowed us to effectively subtract point sources,
small noise features and much of the AGN emission.  Adopting a 5$\sigma$
detection threshold, we obtained the image shown in Figure~\ref{fig:4ims}b.
This clearly shows the north-south bar of emission across the galaxy core,
but also shows a looped structure to the east, and corresponding arms of
emission extending west from the northern and southern ends of the bar.
These correspond to the western cavity and small eastern surface brightness
decrement identified by \citet{Crostonetal08}.

An alternate method for revealing any X--ray structures associated with the
radio lobes is to model the AGN and large scale group halo, subtract these
models from the image and examine any residual features.  Such models need
not be physically meaningful, so long as they provide a good approximation
to the emission components we wish to remove.  We fitted 2-dimensional
surface brightness models using the \textsc{ciao} \textsc{sherpa} package
\citep{Freemanetal01}.  Point sources were removed from the image (as
described in section~\ref{sec:xmm}), and a larger region was used to
exclude the bright X-ray point source RX J1219.8+0545. Models were
convolved with the monoenergetic exposure map and PSF, determined for an
energy of 1 keV, approximating the mean photon energy of the data. A
background image consisting of the scaled particle--only image and scaled
pn OOT events image was used, and convolved and unconvolved flat models
used to account for the X--ray background and any residual particle
background respectively.  

\begin{figure}
\centerline{
\includegraphics[width=\columnwidth, bb=20 200 565 740]{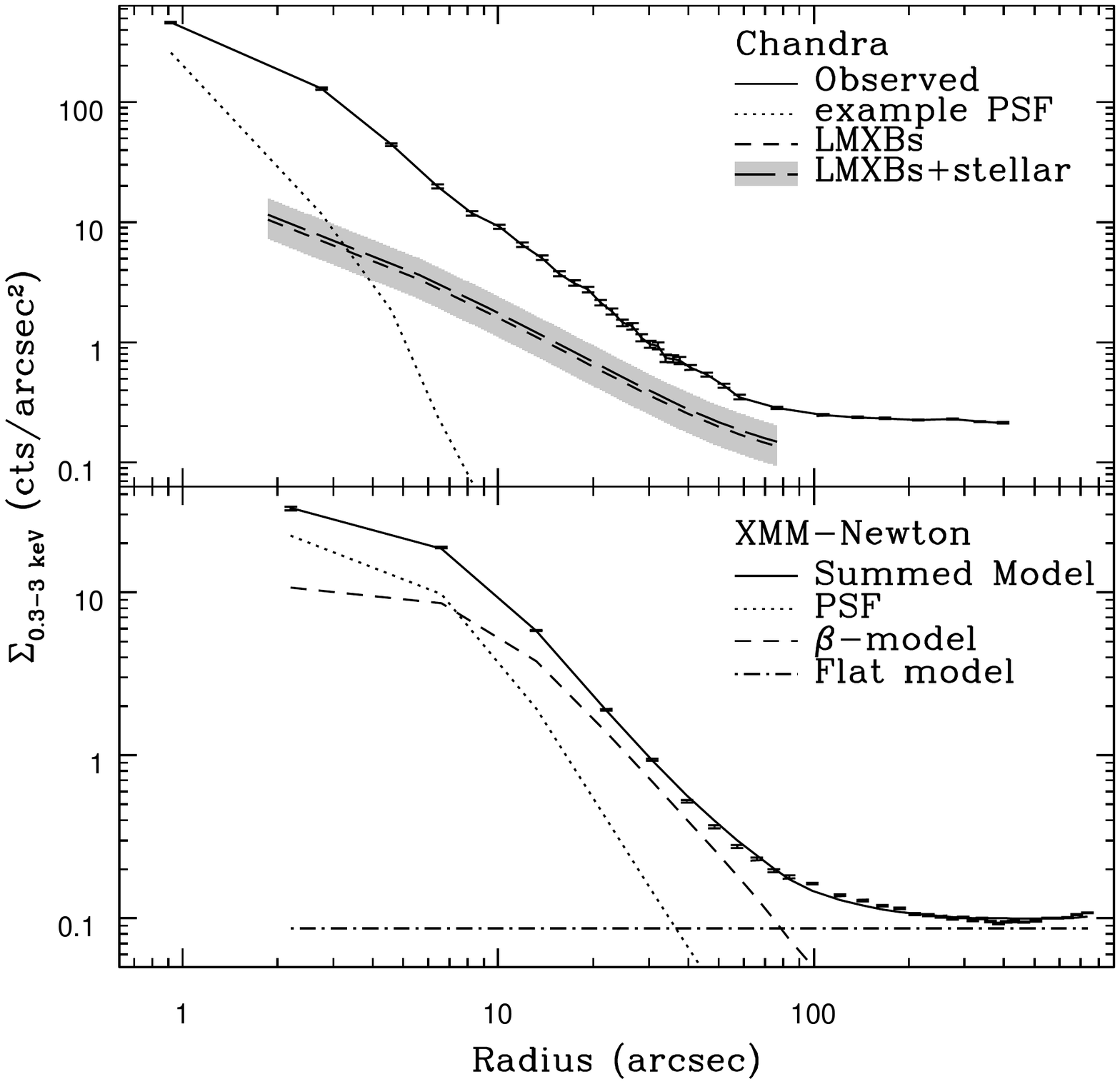}
}
\caption{\label{fig:SB} \textit{Upper panel}: 0.3-3 keV \chandra\ surface
  brightness profile of NGC~4261, compared to estimates of the contribution
  from low-mass X-ray binaries (LMXBs) and stellar sources (coronally
  active binaries and cataclysmic variables) in the galaxy. The observed
  surface brightness is marked by a solid line with error bars indicating
  1$\sigma$ uncertainties. The expected surface brightness from unresolved
  LMXBs and stellar sources is based on the $V$-band surface brightness
  profile of the galaxy, scaled using the relations of
  \citet{KimFabbiano04} and \citet{Sazonovetal06}. The shaded region
  indicates 1$\sigma$ uncertainties on the sum of these two emission
  sources. The dotted line indicates a \chandra\ 1~keV point spread
  function, arbitrarily normalised, for comparison. The 0.3-3~keV emission
  is clearly steeper than the profile expected from stellar sources and
  X--ray binaries (and therefore the optical light profile), but broader
  than the profile expected for a point source.  \textit{Lower panel}:
  XMM-Newton exposure corrected 0.3-3 keV surface brightness profile,
  compared to the fitted model used to search for residual structure.
  Although a five component model was fitted, it is clear that the data are
  primarily described by a central point source, a fairly compact
  $\beta$-model and a flat background.}
\end{figure}

Experimentation showed that modelling the AGN as a point source (a delta
function or narrow gaussian convolved with the PSF) was not effective,
probably because of emission from the jets and dense gas in the centre of
the galaxy. We therefore modelled the core with a $\beta$--model. The model
was fixed to be circular, since the extended emission to north and south
tended to drive the model to extreme ellipticities. The best fitting
parameters for this model were $r_{core}$=1.0$^{+1.2}_{-1.0}$\arcs\ and
$\beta$=0.55$^{+0.01}_{-0.02}$, but we again emphasize that this model is
not intended to be physically meaningful, but simply to provide an
approximation to the core emission, to allow it to be subtracted.  A second
$\beta$--model was added to model large--scale group emission, with slope
and core radius fixed at the values determined from \rosat\
\citep[$r_{core}$=1.68\arcm, $\beta$=0.31][]{Davisetal95}, but a low
normalisation was found for this component, suggesting that the central
$\beta$--model accounted for most of the emission in the field.
Figure~\ref{fig:SB} shows the \xmm\ radial surface brightness profile, with
fitted model components for comparison.

To examine any residual structure, we subtracted the best fitting model
from an image in which the point sources had been removed and replaced
(using the \texttt{dmfilth} task) and the OOT readout streak subtracted.
The resulting residual map is shown in Figures~\ref{fig:4ims}c and d.  The
emission north and south of the core produces the strongest residuals, but
structures associated with the radio lobes are clearly visible. In the
west, arcs of emission extend along the south side of the radio lobe, and
along part of the north side. There is an apparent gap, or weakening of
this emission in the northwest quadrant. On the east, emission
corresponding to the lobe boundary is more diffuse, but extends
approximately to the far end of the radio lobe, and encloses it to north
and south. Agreement with the wavelet smoothed image is generally good,
though the eastern cavity rim is less clearly defined and appears to extend
further east. Immediately outside these structures is a band of negative
residuals, visible as darker regions in Figure~\ref{fig:4ims}c/d.  This is
an artefact of the surface brightness fitting, caused by the model
overestimating the emission at this radius as a consequence of the
north-south bar and cavity rims.

\begin{figure}
\centerline{\includegraphics[width=\columnwidth, bb=20 200 565 740]{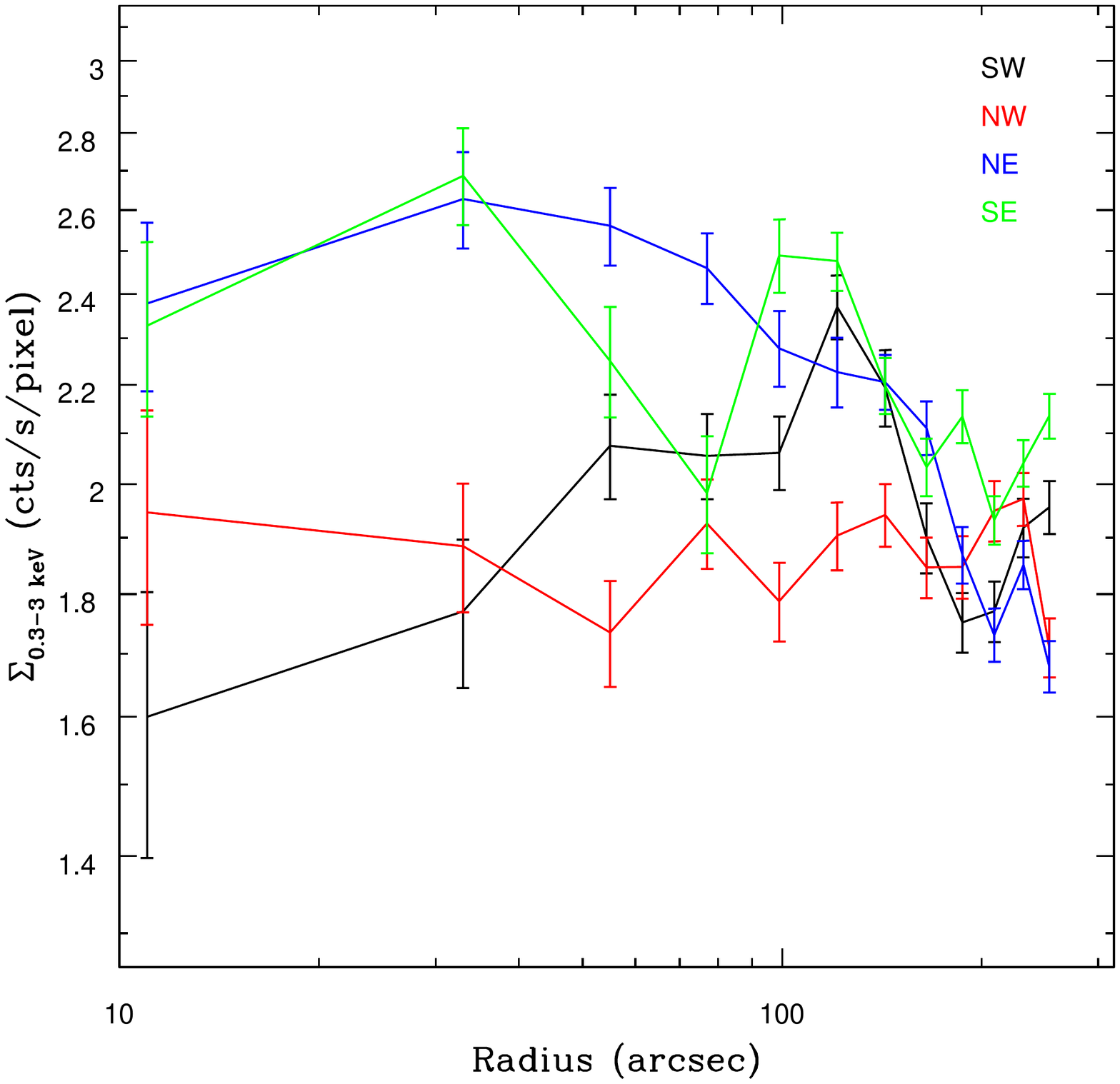}}
\caption{\label{fig:SBarcs} \xmm\ surface brightness profiles across the
  the outer quadrants of the radio lobes, using bins of width 22\arcs\ (5
  EPIC-pn pixels) as shown in Fig.~\ref{fig:4ims}d. The eastern quadrants
  have higher central values, consistent with the residual image, and both
  southern quadrants show a clear peak at $\sim$100\arcs, consistent with
  cavity rims. The northwest quadrant, where a break is observed in the
  residual image, shows no evidence of such a rim. The northeast quadrant
  has a gradual decline from a central peak, probably because both point
  sources and chip gaps affect the area where the rim is expected.}
\end{figure}

To test the significance of these structures we extracted 0.3-3~keV surface
brightness profiles across the eastern and western ends of the lobes. We
used partial circular annuli of width 5 pixels (22\arcs) with limiting
angles chosen to avoid regions which seem to be affected by unsubtracted
point sources. The annuli were centred at R.A. 12$^h$19$^m$14$^s$, Dec.
+05\degree49\arcm53\arcs\ (J2000) for the western and R.A.
12$^h$19$^m$30$^s$, Dec.  +05\degree49\arcm24\arcs\ for the eastern lobe.
The regions, overlaid on the surface brightness residual image to make
their relationship to the rims clear, are shown in Figure~\ref{fig:4ims}d
and the resulting profiles in Figure~\ref{fig:SBarcs}. The difference
between the east and west regions is clear; the western profiles have much
lower central surface brightness than the eastern profiles.  However, both
the southeastern and southwestern profiles show a strong peak at
$\sim$100\arcs, followed by a decline to $\sim$200\arcs, after which the
outer points rise again. The peak at $\sim$100\arcs corresponds to the
structures at the edge of the radio lobes, which are likely cavity rims.
The outer rise appears to be caused by point sources or small clumps of
emission with no clear large-scale structure, or by a large residual to the
southeast surrounding RX J1219.8+0545. We therefore consider the difference
in surface brightness between the highest and lowest values, and find that
the cavity rims are significant at 7$\sigma$ (SW) and 5.7$\sigma$ (SE)
confidence.

The northwestern quadrant shows no significant change in surface brightness
across the edge of the radio lobe, confirming the presence of the break in
the rim seen in this area in the residual image. The northeast profile
shows a slow decline from high values inside the lobe, with no clear rim.
This is probably due to the diffuseness of the emission in this region; the
residual image also shows no clear rim, but a broad area of emission
coincident with the radio and declining towards the end of the lobe.  We
therefore conclude that the lobes of 3C~270 have formed cavities in the
IGM, and that they are surrounded by rims of gas swept up by the lobe
expansion. However, the western rim is either weak or incomplete in its
northwest quadrant, and the complexity of the emission around the eastern
lobe suggests that either the cavity has a complex morphology or that there
are additional emission components present, such as unresolved point
sources.

\subsection{Structure in the galaxy core}
\label{sec:im}
\begin{figure}
\centerline{\includegraphics[width=0.9\columnwidth, viewport=35 170 577 622]{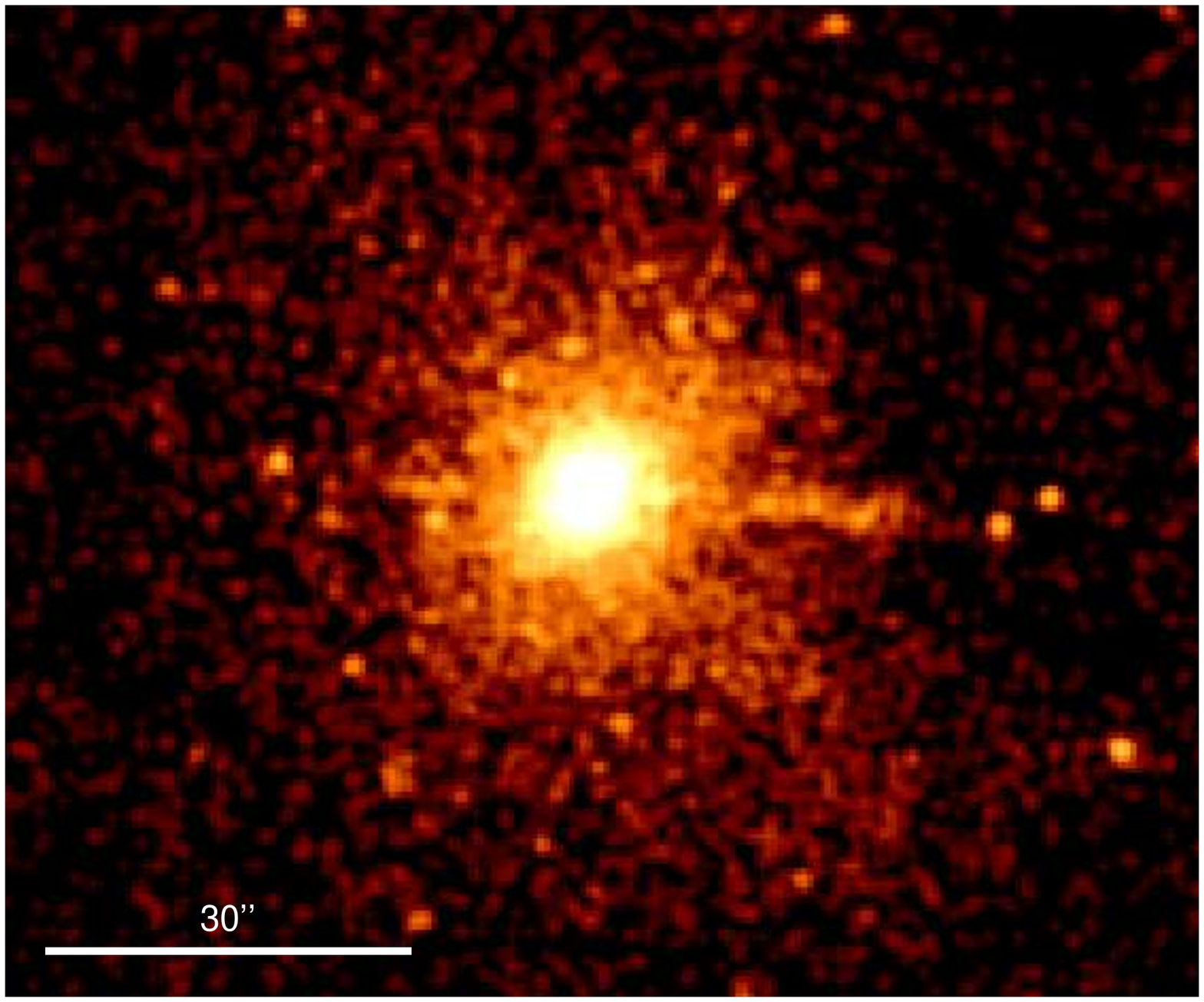}}
\centerline{\includegraphics[width=0.95\columnwidth, viewport=35 170 577 622]{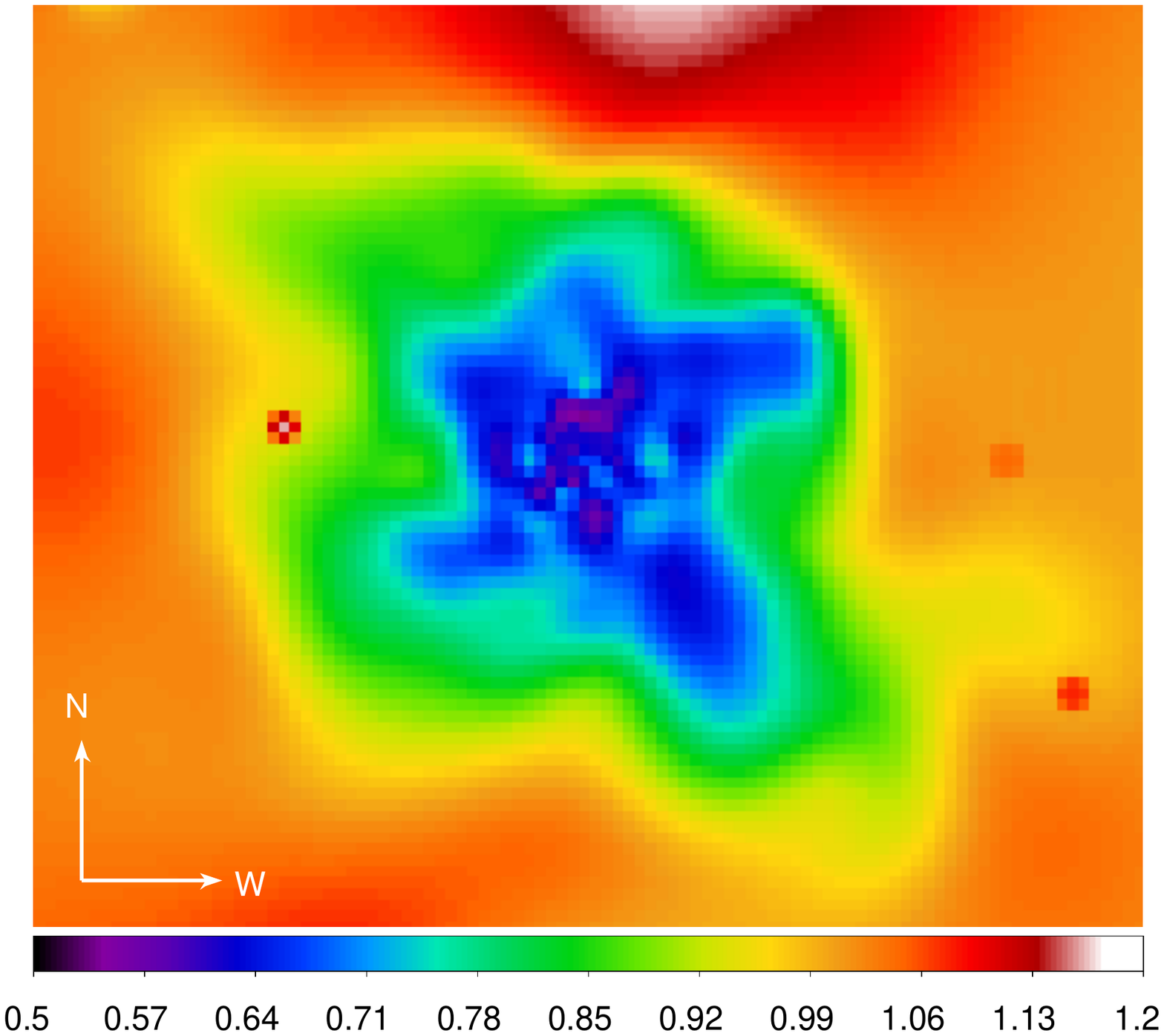}}
\caption{\label{fig:map_LPD} 0.3-2~keV \chandra\ image of the core of
  NGC~4261, with 0.493\arcs\ pixels smoothed with a Gaussian of sigma 2
  pixels (\textit{upper panel}) and adaptively smoothed \chandra\
  temperature map (\textit{lower panel}). The colour bar indicates the
  approximate gas temperature in keV. The two panels have the same
  alignment and angular scale.}
\end{figure}

On smaller scales, we can examine the structure of the diffuse emission
immediately surrounding the AGN and jets.  Figure~\ref{fig:map_LPD} shows a
\chandra\ image of the core and jets. As noted by \citet{Worralletal10},
there are wedge-like surface brightness decrements in the area immediately
north and south of the western X-ray jet, and a suggestion of similar
structure on the eastern side \citep[see also Figs.~2 and 4
of][]{Worralletal10}. The emission bordering these decrements is somewhat
brighter and extends to larger radius than the emission along the
north--south axis, forming an X--shaped structure. The extension to the
southwest is particularly notable, while that to the northwest appears
weakest.  \citet{Worralletal10} suggest that the decrements are caused by
expulsion of the thermal gas by the expanding radio lobes. In this case the
X--structure would likely be this expelled gas, compressed and driven north
and south from the jet axis.

As a simple test of the importance of the different sources of emission in
the core, we compared the 0.3-3~keV \chandra\ surface brightness profile
with that expected from discrete sources in the stellar population of
NGC~4261 (see Figure~\ref{fig:SB}). We estimated the contribution from the
low mass X-ray binary (LMXB) population using a $V$-band optical surface
brightness profile \citep[determined from observations described
in][]{Bonfinietal11} scaled to the expected X-ray luminosity of the LMXB
population, based on the relation of \citet{KimFabbiano04} and taking a
total $K$-band luminosity of 2.2$\times$10$^{11}$\LKsol\ \citep[scaled to
our chosen distance]{EllisOSullivan06}. We note that this is likely to be
an overestimate, since we have already excluded all resolved point sources
but we expect the profile to be largely unchanged. The luminosity is
converted to an expected number of counts in our band using a powerlaw
model with $\Gamma$=1.8. A similar scaling is used to determine the
expected contribution from coronally active binaries and cataclysmic
variables, based on the relation of \citet{Sazonovetal06} and scaling using
a 0.5~keV, solar abundance APEC model. Surface brightness profiles were
determined using elliptical regions chosen to match the optical light
distribution, with axis ratio 1.24 and position angle 68.3\degree\ (where
the position angle is defined as the angle between the major axis and due
west). Regions corresponding to the X-ray jets were excluded, but testing
shows that these make only a minor contribution in the central bins of the
profile. It is clear that the observed surface brightness profile is
steeper than that expected for emission from the stellar population, but
more extended than the \chandra\ 1~keV point spread function, confirming
that much of the emission in this band arises from hot gas.

\subsection{Spectral properties of the cavity rims}
\label{sec:rims}
Given the ongoing jet activity in the system, it is possible that the radio
lobes are still expanding. If they are expanding subsonically, their X--ray
bright rims must consist of compressed gas which may be expected to be
cooling more rapidly than its surroundings. If they are expanding at or
above the sound speed, the lobes will cause shock heating of the
surrounding gas. Using \xmm, we found no significant temperature
differences between the cavity rims, and regions immediately inside and
outside the cavities. This is unsurprising, since the regions used are
necessarily large and include unresolved point sources and gas at a range
of temperatures.

\begin{figure}
\centerline{\includegraphics[width=\columnwidth, bb=20 200 565 740]{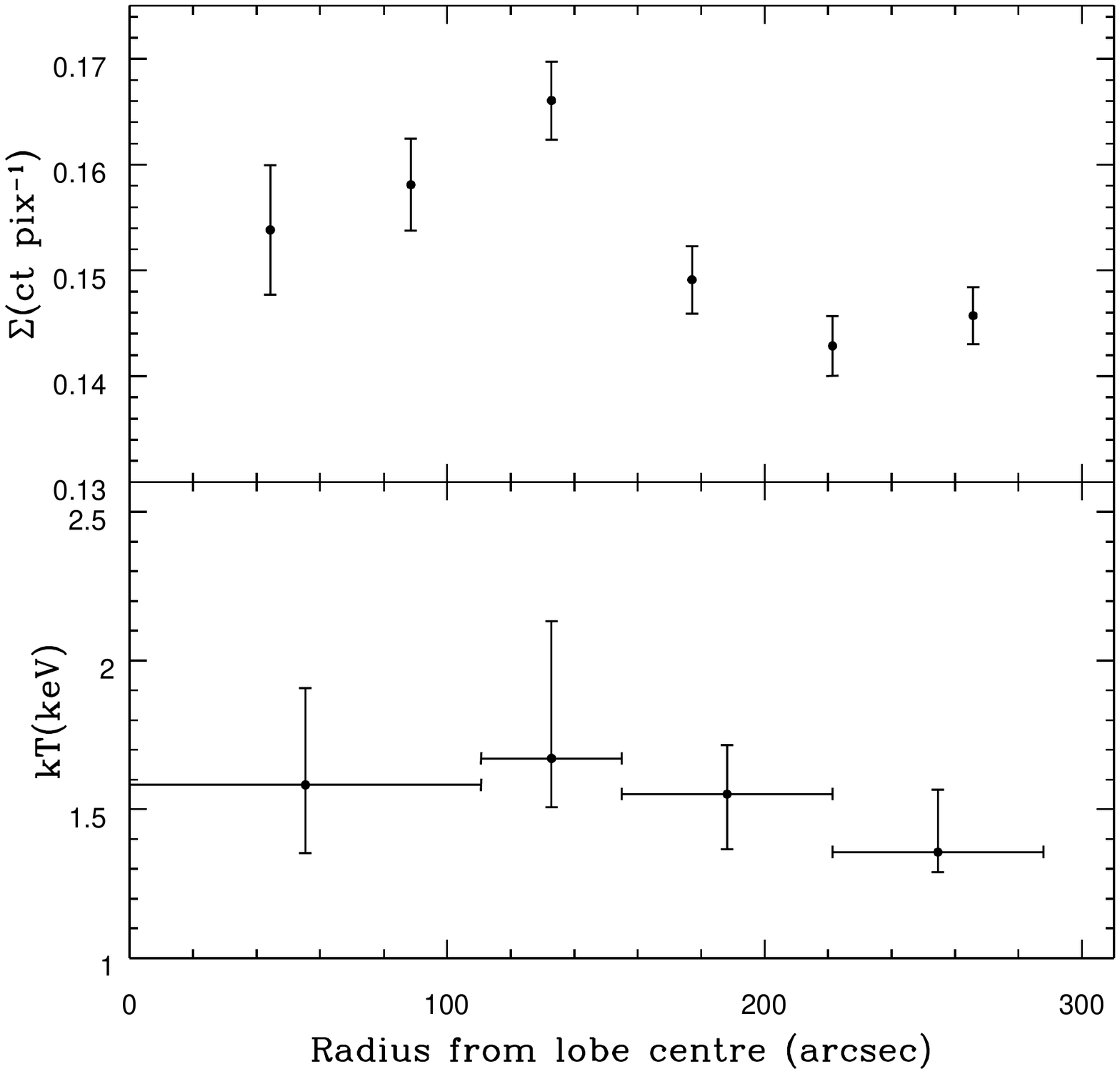}}
\caption{\label{fig:chanSB} 0.3-3~keV Surface brightness and temperature
  from elliptical annuli extending across the southern edge of the eastern
  radio lobe. Radii are measured along the minor axis of the ellipses,
  which is aligned roughly south. The rim of the lobe falls in bin 3 of the
  surface brightness profile.  Surface brightness uncertainties are
  1$\sigma$, temperature uncertainties 90\%}
\end{figure}

While the field of view of the ACIS S3 CCD limits our ability to examine
the edges of the radio lobes with \chandra, we can examine the southern
parts. Figure~\ref{fig:4ims}a shows a series of partial elliptical annuli
which were used to search for a surface brightness jump across the
southeast lobe edge. These regions were chosen to match the shape of the
apparent surface brightness feature, with large widths to increase the
signal-to-noise ratio in each bin. Figure ~\ref{fig:chanSB} shows the
exposure corrected surface brightness in each bin. A peak is found at the
position of the lobe rim, where the surface brightness exceeds that
immediately outside the rim at $>$3$\sigma$ significance (a 3.4$\sigma$
difference between bins 3 and 4).  This corresponds reasonably well with
the feature observed in the \xmm\ images.

If this surface brightness increase is the result of a shock, the standard
Rankine-Hugoniot jump conditions \citep[e.g.,][]{LandauLifshitz59} can be
used to estimate the shock velocity.  The change in surface brightness of a
factor $\sim$1.15 indicates a shock with Mach number
$\mathcal{M}=1.05\pm0.02$, suggesting that if a shock is present, the lobe
is expanding only marginally supersonically. The presence of a shock can
only be confirmed by the detection of a temperature increase. Spectra were
extracted across the rim, using larger regions inside and outside to
maximise the signal-to-noise ratio. The spectra contained $\sim$1900-5400
counts, with the background contributing 24-35 per cent.
Figure~\ref{fig:chanSB} shows the temperatures measured from an absorbed
APEC model fitted to these spectra, and the fit parameters are shown in
Table~\ref{tab:rims}. The uncertainties are large, as expected for
relatively small regions with extensive foreground and background group
emission, and while the best fitting temperature of the rim is higher than
that of its surroundings, the temperature difference is not statistically
significant. However, the temperature increase expected for a
$\mathcal{M}=1.05$ shock is only a factor 1.05, and the data are consistent
with heating at this level.

\begin{table*}
\begin{center}
\caption{\label{tab:rims} Fit parameters for the \chandra\ spectra
  extracted across the southern sector of the east lobe.}
\begin{tabular}{lccccc}
Model & Parameter & \multicolumn{4}{c}{Bin} \\
      &           & 1 & 2 & 3 & 4 \\
\hline\\[-2.5mm]
wabs*apec & $kT$ (keV) & 1.58$^{+0.33}_{-0.23}$ & 1.67$^{+0.46}_{-0.16}$ & 1.55$^{+0.17}_{-0.18}$ & 1.36$^{+0.21}_{-0.07}$ \\
          & Abundance (\Zsol) & 0.67$^{+0.95}_{-0.29}$ & 0.85$^{+1.06}_{-0.33}$ & 0.57$^{+0.56}_{-0.22}$ & 0.50$^{+0.35}_{-0.16}$ \\
          & $\chi^2$ / d.o.f. & 68.56 / 84 & 79.9 / 74 & 172.54 / 156 & 175.23 / 189 \\[+2mm]
wabs*(apec+powerlaw) & $kT$ (keV) & 1.65$^{+0.39}_{-0.24}$ & 1.70$^{+0.46}_{-0.20}$ & 1.54$^{+0.16}_{-0.21}$ & 1.35$^{+0.19}_{-0.07}$ \\
with $\Gamma$=2\         & Abundance (\Zsol) & 1.40$^{+1.43}_{-0.90}$ & 4.74$^{+2.3}_{-4.74}$ & 0.83$^{+0.56}_{-0.51}$ & 0.53$^{+0.35}_{-0.18}$ \\
                     & $\chi^2$ / d.o.f. & 68.26 / 83 & 78.94 / 73 & 172.09 / 155 & 175.23 / 188 \\[+2mm]
wabs*powerlaw        & $\Gamma$ & 2.06$^{+0.31}_{-0.28}$ & 1.87$^{+0.33}_{-0.30}$ & 2.17$^{+0.22}_{-0.21}$ & 2.15$^{+0.19}_{-0.18}$ \\
                     & $\chi^2$ / d.o.f. & 120.25 / 85 & 122.29 / 75 & 234.52 / 157 & 274.43 / 190 \\
\end{tabular}
\end{center}
\medskip
The powerlaw photon index was fixed at $\Gamma$=2 in the apec+powerlaw
model, as it was poorly constrained when freely fitted. 
\end{table*}

\chandra\ observations of the southwest radio lobe of Centaurus A have
shown that the X--ray emission at the boundary of the lobe is best
characterised as a single component absorbed powerlaw model, and arises in
large part from synchrotron emission \citep{Crostonetal09}. To test the
possibility of a synchrotron component contributing to the emission around
the lobes of NGC~4261, `powerlaw' and `apec+powerlaw' models were fitted to
the spectra extracted across the lobe rim. Fitted model parameters are
given in Table~\ref{tab:rims}. An absorbed powerlaw model was a poor fit to
all spectra. The addition of a powerlaw to the absorbed apec models did not
significantly improve the quality of fits, and in all cases the photon
index of the powerlaw was poorly constrained and the normalisation
consistent with zero.  Fixing the photon index at $\Gamma$=2 (compatible
with the data and similar to the values found for Centaurus A), we are able
to place a 90 per cent.  upper limit on the synchrotron flux from this
portion of the lobe rim of F$_{\rm
  X,0.3-7.0}\le$3.18$\times$10$^{-14}$\ergpspcmsq. Scaling to the total
area of the lobe rims, this is equivalent to an upper limit on luminosity
of L$_{\rm X,0.3-7.0}\la$3.3$\times$10$^{40}$\ergps. The equivalent
(detected) thermal luminosity of the rims is L$_{\rm
  X,0.3-7.0}$=4.15$\times$10$^{40}$\ergps. A similar analysis of the
north-south bar also found no evidence of non-thermal emission.

The \chandra\ profiles extend across the cavity rim in a direction roughly
perpendicular to the jet axis. We might expect a higher expansion rate, and
therefore greater surface brightness differences (and potentially
temperature differences), at the jet tips.  The lack of a detected rim
corresponding to the northwest quadrant of the western radio lobe (see
Section~\ref{sec:im2}) is informative in this regard, and we do not see
strong surface brightness features corresponding to the east or west
extremes of either lobe.  However, if the expansion were very subsonic, gas
swept up by the lobes would be able to flow out into regions of lower
pressure, and a rim feature would be weak or nonexistent. It is therefore
likely that the lobe expansion rate is comparable to the external sound
speed. We conclude that the cavity rims consist of material swept up and
compressed by the expansion of the lobes, and at most only weakly heated by
shocks or compression.

\subsection{Spectral map}
To examine the spatial variation of temperature in the gas, we prepared a
temperature map using the technique developed by \cite{Davidetal09} which
takes advantage of the close correlation between the strength of lines in
the Fe-L complex and gas temperature in $\sim$1~keV plasma.  Most of the
emission from such gas arises from the L--shell lines from Fe-XIX
(Ne--like) to Fe-XXIV (He--like). For CCD resolution spectra, these lines
are blended to form a single broad peak between approximately 0.7 and 1.2
keV. The centroid or mean photon energy of this peak increases with the
temperature of the gas as the dominant ionisation state of Fe shifts from
Fe XIX in 0.5~keV gas to Fe XXIV in 1.2 keV gas, with Fe XVII and Fe XVIII
providing the strongest line emission at the temperatures seen in the core
of NGC~4261. The mean photon energy of the blended L--shell lines is
independent of energy above $\sim$1.2~keV.

We can thus estimate the temperature distribution of the gas by mapping the
mean photon energy in the 0.7-1.2~keV band. To determine the relation
between $kT$ and mean photon energy we require information on the detector
response and the typical properties of the gas. We therefore extracted a
spectrum from a $\sim$1\arcm\ region centred on the galaxy, excluding the
central point source and jets, and fitted it using an absorbed apec model.
Based on this fit, we chose a value of 0.5\Zsol as representative of the
abundance and temperature of the central part of the group, and simulated
spectra based on this model, with redshift set to that of NGC~4261 and the
hydrogen column set to the galactic value.  These showed that for
temperatures between $\sim$0.5 and $\sim$1.1~keV, the relationship is
approximately linear ($kT=-4.43+5.68<E>$). Prior testing in other systems
suggests that the relationship is relatively insensitive to variations in
abundance and hydrogen column \citep{OSullivanetal09}.

Previous studies have shown the core temperature of NGC~4261 to be
0.6-0.7~keV \citep[e.g.,][]{Worralletal10} and the IGM temperature to be
$\sim$1.6~keV \citep[e.g.,][]{Humphreyetal09b}. The core is therefore
suitable for mapping using this technique. The lower panel of
Figure~\ref{fig:map_LPD} shows the resulting map. The estimated
temperatures are comparable with those derived from previous spectral fits
(and our own radial temperature profile, see Section~\ref{sec:spec}).

We do not expect to see the jet in the temperature map, since its powerlaw
emission is spectrally flat compared to the strongly peaked plasma
component, and indeed no features corresponding to the jet are observed in
the map. The X--structure observed in surface brightness is clearly
visible, which confirms that the core gas distribution is disturbed and
that the arms of the X consist of cool gas with a temperature
$\sim$0.65~keV. The regions of low surface brightness along the west jet
have a higher temperature (0.8-0.9~keV), in agreement with the temperature
measured by \citet{Worralletal10}. On slightly larger scales gas with
temperatures $\la$0.95~keV appears most extended to the northeast and
southwest. The cavity rims and the north-south bar are not observed in the
temperature map, confirming that these do not contain significant
quantities of cool gas. We conclude that NGC~4261 hosts a small cool core,
whose structure has been disturbed by the expansion of the radio lobes and
which is at present not spherically symmetric.

\subsection{Radial Spectral Analysis}
\label{sec:spec}
\begin{figure}
\centerline{\includegraphics[width=\columnwidth, bb=20 200 565 740]{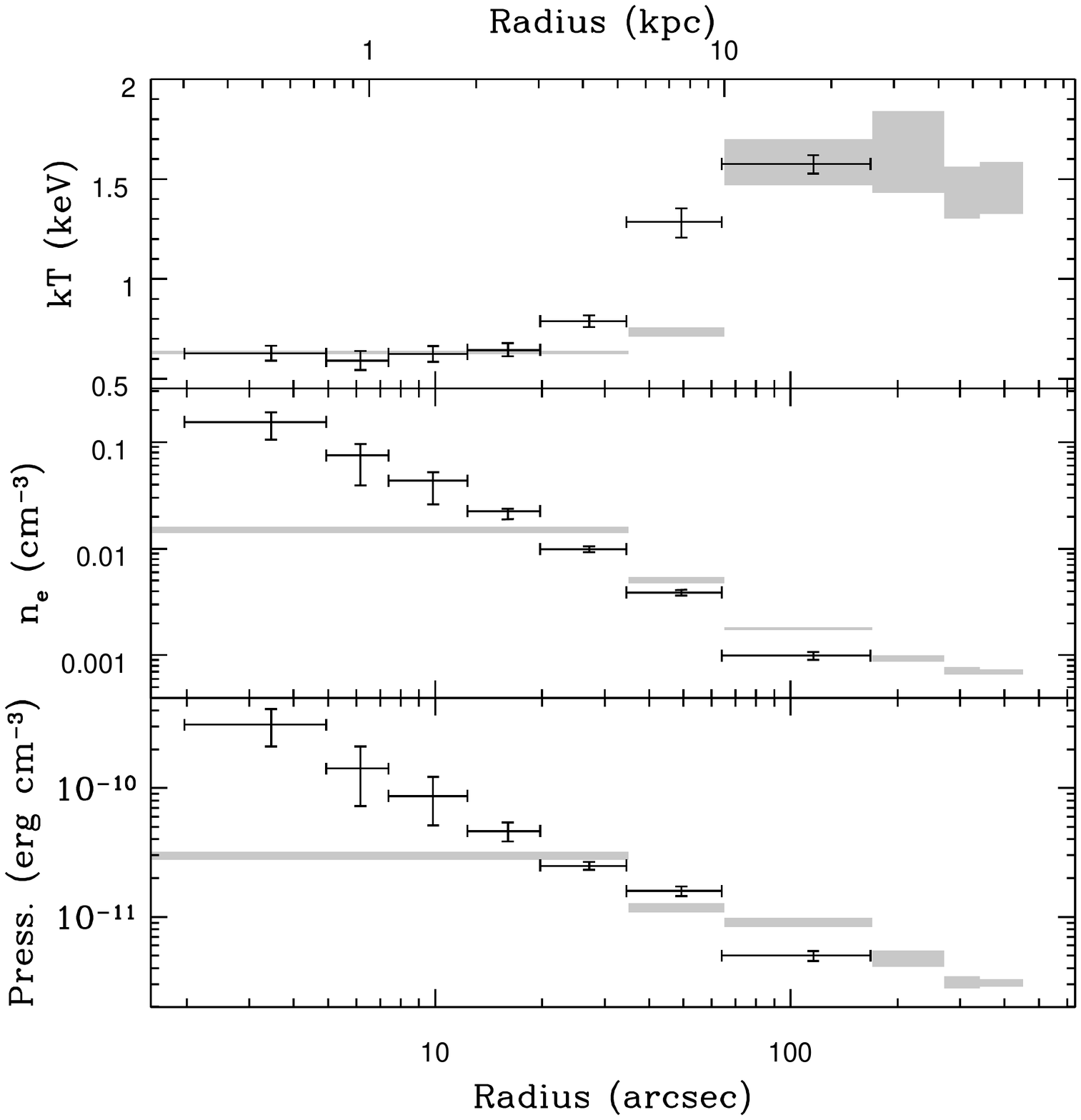}}
\caption{\label{fig:deproj} Deprojected temperature, density and pressure
  profiles, derived from the ACIS S3 (errorbars) and \xmm\ (grey
  rectangles) observations.}
\end{figure}

In order to examine the underlying structure of the intra-group medium and
estimate the energy required to form the cavities, we extract spectral profiles
from the \chandra\ and \xmm\ data. Regions corresponding to the cavities
are excluded, and the core and jet are excluded from the \chandra\ spectra
using a rectangular region of length 30\arcs\ and width 4\arcs\ centred on
the nucleus. For the \chandra\ data, annular regions were selected to
contain 3000 net counts, and for \xmm\ a total of 10000 net counts across
the three EPIC cameras. The only exception is the innermost bin of the
\xmm\ profile, whose radius was fixed at 35\arcs\ so as to simplify
comparison with the \chandra\ profile. The spectra were fitted with a
deprojected, absorbed APEC model. An additional powerlaw component was
included to account for emission from the AGN in the innermost bin, and
from unresolved point sources in the central three \xmm\ bins. Abundances
were tied between bins where necessary to stabilise the deprojection.
Figure~\ref{fig:deproj} shows the resulting temperature and density
profiles, and the pressure profile derived from them. Pressure was
calculated as $P=nkT$ where $n=2n_e$.

The \chandra\ and \xmm\ profiles agree reasonably well at large and small
radii, but there is some disagreement in temperature in the
$\sim$35-65\arcs\ bin, and in density between $\sim$35\arcs\ and 170\arcs.
The temperature disagreement is likely caused by the difference in the
radii of the outermost bins of the two profiles, and the relatively low
resolution of the profiles in comparison to the temperature gradient,
particularly in the \xmm\ profile, where the full range of temperatures
across the gradient is represented by only a single model temperature in
the deprojection. Scattering of low energy emission outward from the
central bin is also a possibility, given the broad point spread function of
\xmms\ and the high surface brightness of the core.  However, the large
spectral extraction region of the central bin should reduce this effect.
The density difference may be a product of the difference in extraction
regions, since the \xmm\ annuli cover a larger area north of the core than
the ACIS spectra, which reach the edge of the S3 CCD. The outermost annuli
of both profiles are excluded since they contain emission from the outer
parts of the group halo (which extends beyond the field of view) and would
thus give incorrect density estimates.  However, the level of agreement is
sufficient for our purposes.  We note that the \chandra\ profiles also
agree well with those of \citet{Humphreyetal09b}.

The shapes of the temperature and density profiles are typical of cool-core
galaxy groups. Defining the cooling region as that volume within which the
temperature declines ($\sim$10~kpc radius), the cool core is relatively
small. The core has a relatively constant temperature of $\sim$0.6~keV, in
agreement with the estimate in the temperature map. The temperature peaks
at $\sim$1.6~keV, between 10 and 40~kpc from the AGN, and flattens or
declines at larger radii. The pressure profile appears relatively smooth,
though this may be misleading since the annular average conceals azimuthal
variations (c.f., Figure~\ref{fig:map_LPD}).

\section{Analysis}
\label{sec:ana}

\subsection{Outburst energy budget and timescale}
\label{sec:energy}
To estimate the mechanical energy output of the AGN, we approximate the
area of the cavities with ellipses, and assume rotational symmetry about
their major axis. The ellipses are shown in Figure~\ref{fig:4ims}c. Neither
the shape of the radio lobes nor the shape of the cavities in the X-ray is
straightforward, and therefore there is significant uncertainty in the
following estimates. The east cavity in particular is poorly defined, and
could be smaller than we have assumed. The lobes may also be overpressured
and still expanding, which would lead to an underestimate of the energy they
contain.

Assuming the lobes to be completely filled by relativistic plasma, we can
estimate their enthalpy to be 4$PV$ where $V$ is the lobe volume and $P$
the external pressure. The two ellipses are defined as having semi--major
and --minor axes of 18.3$\times$16.7~kpc (West) and 20.7$\times$15.3~kpc
(East). Their size is such that the external thermal pressure of the IGM
varies from $\sim$3-80$\times$10$^{-12}$\ergpcmcu along their length.  We
use pressure at the midpoint of the lobes, from the \chandra\ profile,
$\sim$5$\times$10$^{-12}$\ergpcmcu, but note that the \xmm\ pressure
estimate is a factor of 2 higher. We therefore estimate the enthalpy of the
two cavities to be $\sim$2.4$\times$10$^{58}$~erg.

The timescale over which cavities have formed, and therefore of the
associated AGN outburst, is often estimated from dynamical arguments
\citep[e.g.,][]{Churazovetal01, Dunnetal05}. In this system, the lobes
appear still to be connected to the core, and their shape suggests that the
jets are still the dominant factor determining the lobe size and location,
rather than buoyant forces. We can estimate the timescale of formation
based on the time taken for the lobes to expand to their current size at
the external sound speed, but as it is likely that the lobes expanded more
rapidly in the past, and are only now approaching pressure balance with the
surrounding IGM, this should be considered as an upper limit on the
timescale of formation. For $kT$=1.5~keV, the sound speed is 585\kmps. The
sound speed in the cool central region is considerably lower (380\kmps) but
it seems likely that expansion through this region will have happened at
the onset of the outburst, when the source is likely to have been strongly
overpressured.  The tips of the lobes are $\sim$38~kpc (West) and 45~kpc
(East) from the central point source, suggesting that the lobes must be at
most 75~Myr old. The mechanical power output of the AGN (including both
jets) is therefore $\geq$10$^{43}$\ergps, with greater power output
indicated if the radio lobes expanded supersonically over some portion of
their history. This value is consistent with the energy output estimated
for NGC~4261 by \citet{Crostonetal08}. It exceeds the estimate of
\citet{Cavagnoloetal10} by an order of magnitude, but Cavagnolo et al.  did
not identify the main cavities, considering only the structures associated
with the jets in the core region. Lower limits to the total radiative power
output of the AGN can be taken from the intrinsic X--ray luminosity of the
nucleus \citep[1.4$\times$10$^{41}$\ergps][]{Worralletal10}, or the
extended radio luminosity at 178~MHz \citep[$\nu$L$_\nu$=10$^{40}$\ergps or
L$_\nu$=24.73 W~Hz$^{-1}$][]{Balmaverdeetal06,Chiabergeetal99}.

The longer eastern lobe extends to $\sim$5\arcm\ ($\sim$45~kpc) from the
nucleus.  The total X--ray luminosity of the gas within this radius is
$\sim$1.6$\times10^{41}$\ergps (0.3-7.0~keV), of which approximately two
thirds arises in the central 10~kpc. The mechanical power output of the AGN
exceeds this value by at least a factor of 60. If the formation timescale
of the cavities is longer than we have estimated, this discrepancy will be
reduced, but given the evidence for near--sonic expansion of the lobes, it
is unlikely that the radiative losses can balance the energy output of the
radio source.  Even if we consider only the work done in moving the thermal
gas as the lobes expand, $PV$, the energy involved would certainly be
sufficient to balance cooling.

\subsection{Energy balance in the cool core}
\label{sec:CC}
We consider the state of the cool core of NGC~4261 by examining the energy
balance between radiative cooling and heating. Heating could be supplied by
the AGN, by supernovae in the core, or through conduction from the hotter
surrounding IGM. We define the core as including the inner six bins of the
\chandra\ deprojected temperature profile, giving an outer radius of
$\sim$9.8~kpc ($\sim$1\arcm), but note that \citet{Worralletal10} have
estimated that up to $\sim$20 per cent. of the gas in this volume may have
been evacuated by the radio source. As discussed in
Section~\ref{sec:energy}, the mechanical power of the jets is
$\geq$10$^{43}$\ergps, while the total X--ray luminosity of the gas in this
region is $\sim$10$^{41}$\ergps (0.3-7.0~keV). The jets therefore need only
heat the gas with 1 per cent. efficiency to balance cooling.  Taking a
maximum temperature of 1.6~keV for the IGM and a typical core temperature
of 0.6~keV, we can estimate the energy that would be required to heat the
cool core to the temperature of the IGM, $\sim$9.1$\times$10$^{56}$~erg.
Given the timescale over which the AGN has been active (75~Myr at most),
heating with an efficiency of 3.5 per cent would have destroyed the cool
core. In fact, the heating efficiency is likely to be considerably lower,
since the core still has $kT$=0.6~keV and the mechanical power of the jets
will be greater if the timescale of the outburst is shorter than our
estimated limit.

Given the temperature gradient between the core and surrounding IGM,
conduction could potentially be a significant heat source. Following the
methods described in \citet{OSullivanetal07}, and working from the
properties of the IGM gas immediately outside the core (bin 7 of the
\chandra\ temperature and density profiles), we estimate the mean free path
of electrons (in the absence of significant magnetic field) to be
$\sim$8~kpc. Since this is comparable to the scale of the temperature
gradient, conduction into the cool core is likely to be saturated
\citep{CowieMcKee77}. The saturated heat flux per unit area into the core
will be:

\begin{equation}
\noindent q_{sat} = m_e\left(\frac{2kT}{\pi m_e}\right)^{1/2}n_ekT \,\,\,\,\,\,\,\,\ergpspcmsq.
\label{eqn:cond}
\end{equation}

Assuming the core to be spherical, the saturated conduction rate would thus
be 1.8$\times$10$^{42}$\ergps. Given the highly aspherical structure of the
core, the surface area over which conduction can occur will be higher,
leading to an increased rate. However, this estimate does not account for
the effect of magnetic fields and should therefore be considered an upper
limit. While a field alignment which increases conduction is possible, it
is unlikely and the common assumption of a tangled field would reduce
conduction significantly. We discuss this further below.

While a factor of $\sim$6 less than the mechanical energy available from
the jets, the conduction rate still exceeds the radiative cooling rate by
a factor of $\sim$18, suggesting that conduction must be strongly
suppressed for the gas to have retained its cool temperature. Heating at
this conduction rate would destroy the cool core in 16~Myr, a shorter
timescale than that required to form the cavities. Assuming a timescale
equal to that of the AGN outburst, we find that conduction must be
suppressed below the rate estimated from equation~\ref{eqn:cond} by a
factor of $>$5.

However, this timescale is almost certainly too short. There are several
reasons to believe that the temperature structure of the IGM prior to the
AGN outburst must have been similar to that currently observed. Firstly,
most nearby galaxy groups have a temperature structure similar to that of
NGC~4261, with a central cool core, peak in temperature at moderate radii
and temperature decline at large radii \citep{RasmussenPonman07,Sunetal09},
indicating that this is probably a moderately stable state.  Secondly, the
formation of the observed temperature gradient over a timescale comparable
to that of the outburst would be difficult. If the AGN formed the
temperature gradient by heating the ICM at 10-70~kpc, the energy required
would be $\sim$7.5$\times$10$^{58}$~erg (assuming an initial temperature
equal to that of the core, $\sim$0.6 keV).  This exceeds the enthalpy of
the cavities by a factor of 3, and would require strong additional heating
of the gas, presumably by shocks, for which there is no evidence. Since the
virial temperature of the group would be $\sim$0.6~keV in this scenario,
such strong heating would be likely to cause convective motions in the IGM,
and might be sufficient to unbind gas from the group gravitational
potential. The heated region also extends well outside the tips of the
radio lobes (70~kpc compared to $\sim$45~kpc), but the timescale over which
heated gas could disperse outward is longer than the AGN outburst
timescale. The sonic timescale for gas in the hot region corresponding to a
distance of 60~kpc is $\sim$100~Myr, and convective motions would take
several times longer to transport the gas outward. It is therefore likely
that the appropriate timescale over which to consider conduction into the
cool core is much longer than that of the AGN outburst, and the suppression
factor for conduction is correspondingly higher.

A suppressed conduction rate is not unexpected in the cool core of
NGC~4261. The simplest assumption of a tangled magnetic field would lead to
significant suppression. Magnetic field lines aligned radially along a
temperature gradient, which would enhance conduction, are likely to be
realigned by convective gas motions, taking on a tangential alignment which
will strongly suppress conduction. This heat-flux-driven buoyancy
instability \citep[HBI,][]{ParrishQuataert08} can reduce the conduction
rate by large factors. The timescale for saturation of the instability is
long and dependent on the steepness of the temperature gradient
\citep{Parrishetal09}, and any field alignment can be disrupted if
turbulent motions are introduced into the gas \citep{Parrishetal10}.
However, HBI provides a potential mechanism for suppressing conduction to
the level required.

The number of supernovae occurring within the core can be estimated based
on the stellar population in the region. NGC~4261 was taken to be a prolate
ellipsoid whose major axis lies in the plane of the sky, and which is
described by a \citet{Sersic68} model with effective radius 56.81\arcs,
axis ratio 0.80, and shape parameter n=5.44 \citep[determined from
observations described in][]{Bonfinietal11}. These physical parameters are
consistent with the limits determined from stellar velocity measurements,
ignoring the probable small inclination of the galaxy away from the plane
of the sky \citep{DaviesBirkinshaw86}. The S\'{e}rsic model was deprojected
using the three dimensional luminosity density distribution described by
\citet{ChakrabartyJackson09}, normalising to the total optical luminosity
\LB=5$\times10^{10}$\LBsol.

Assuming a supernova rate of 0.166 per 100~yr per 10$^{10}$\LBsol\
\citep{Cappellaroetal99}, and an energy release of 10$^{51}$~erg, we
estimate the heating rate within the core to be 6.4$\times$10$^{40}$\ergps,
if the energy release is efficiently coupled to the gas. This agrees within
a factor of 2 with the X--ray luminosity of the core.  However, since the
X--ray surface brightness profile is considerably more peaked than the
optical light profile (see Fig.~\ref{fig:SB}), it seems likely that supernova heating will be most
effective in the outskirts of the cooling region, where it may balance
radiative cooling, but that it will be unable to prevent cooling in the
high density gas near the AGN.

To determine whether the cool core could potentially fuel the AGN, we
estimate the isobaric cooling time and mass deposition rate in the
innermost \chandra\ spectral bin (radius $\sim$0.3-0.8~kpc). We find a
cooling time of $\sim$60~Myr, and a mass deposition rate
$\dot{M}_{cool}$=0.033\Msol\ yr$^{-1}$. Comparing this with our estimate of
the mechanical power output of the AGN, we find that the efficiency of the
AGN in converting cooling gas into energy would be
$e_{conv}=P_{mech}/\dot{M}c^2$=1.8$\times$10$^{-3}$. \citet{Zezasetal05}
argued for an accretion rate much lower than this mass deposition rate
($4.3\times10^{-6}\eta^{-1}$\Msol~yr$^{-1}$, based on the Eddington ratio
of the AGN) but it is unclear whether the core X--ray emission component
used to make this estimate arises from radiatively inefficient accretion,
or from the radio jets \citep{Zezasetal05,Worralletal10}. Based on the rate
of stellar mass loss from AGB stars \citep{Atheyetal02}, we estimate that
$\sim$0.095\Msol\ yr$^{-1}$ of gas is injected into the core from stars.
Supernovae will make only a minor mass contribution, based on the rate of
$2\times10^{-3}$~SN~yr$^{-1}$, but will contribute to heating the gas to
the observed temperature of $\sim$0.6~keV. The total mass injection rate is
similar (within the large uncertainties) to the rate of loss through
cooling, suggesting that there is no need for a flow of cooling gas into
the core to sustain its current size and gas content. It therefore seems
possible for the cool core to provide cooling gas to fuel the AGN over long
timescales, replenished only by gas lost from the stellar population within
the core.

In summary, it seems likely that neither the AGN jets nor conduction from
the surrounding IGM is effective in heating the cool core. Supernova
heating may balance radiative cooling in the outer part of the core, but is
unlikely to do so in the denser inner regions.  However, stellar mass
losses within the core should be sufficient to replenish any gas lost,
allowing the core to maintain its current size and to continue to supply
cooling gas to fuel AGN activity over timescales longer than that of the
current AGN outburst.

\subsection{Pressure balance in the lobes}
\label{sec:lobes}
As discussed in Section~\ref{sec:rims}, while the lobes may be
overpressured with respect to their environment, the difference in
pressures is probably not large. Given approximate pressure equilibrium, a
comparison of the apparent pressures of the thermal plasma of the IGM and
the relativistic plasma of the radio lobes can provide insight into the
particle content of the lobes
\citep[e.g.,][]{HardcastleWorrall00,Dunnetal05,Birzanetal08}. The lobes of
FR\,I radio sources are generally found to have apparent synchrotron
pressures well below the IGM pressure \citep[e.g.,][]{Ferettietal92}, and
this is generally assumed to indicate either that the lobes contain a
significant population of non-radiating particles, or that the assumption
of equipartition is incorrect, with the magnetic field providing additional
pressure support.

We estimate the physical properties of the radio lobes from the 1.5~GHz VLA
archival data. The volume of the lobes is derived from the ellipse regions
in Figure~\ref{fig:4ims}, as described in Section~\ref{sec:energy}.  We
adopt an electron spectral index of p=2.2, equivalent to a radio spectral
index of $\alpha$=0.6, which compares well with the flux density
measurements compiled in NED\footnote{The NASA/IPAC Extragalactic Database
  (NED) is operated by the Jet Propulsion Laboratory, California Institute
  of Technology, under contract with the National Aeronautics and Space
  Administration.}, and the value of $\alpha$=0.62$\pm$0.17 estimated by
\citet{Kuhretal81}. We make the usual assumptions of minimum energy
conditions, in which the total energy content of the relativistic particles
and magnetic field are roughly equal, and a low energy cutoff in the
electron energy distribution at a Lorentz factor $\gamma_{min}=10$. We can
then estimate the strength of the minimum energy magnetic field $B_{min}$
and the pressure of the relativistic plasma P$_{min.en.}$, which is defined
to be

\begin{equation}
\noindent {\rm P}_{min.en.} = \frac{B_{min}^{2}}{2\mu_0\epsilon}+\frac{(1+{\rm k})E_e}{3V\phi},
\label{eqn:psync}
\end{equation}

\noindent where $V$ is the lobe volume, $E_e$ is the energy of the electron
population and $\mu_0$ is the permeability of free space
(4$\pi\times10^{-7}$ in S.I. or 4$\pi$ in $cgs$ units). $\phi$ is the
filling factor of the lobes (assumed to be 1) and k the ratio of energy in
non-radiating particles to the energy in relativistic electrons (assumed to
be 0). The estimated parameters, assuming (1+k)/$\phi$=1, are shown in
Table~\ref{tab:radio}. The factor $\epsilon$ is dependent on the ordering
of the magnetic field, with $\epsilon$=3 for a tangled field, and
$\epsilon$=1 for a uniform or stretched field \citep{Leahy91}. We have
adopted $\epsilon$=3, but we note that $\epsilon$=1 is assumed in some
commonly used formulae \citep[e.g.][]{Burnsetal79,Fabianetal02}. For
comparison, adoption of $\epsilon$=1 would approximately double our
estimate of lobe pressure.

\begin{table}
\caption{\label{tab:radio}Magnetic field, pressure and particle content of the radio lobes} 
\begin{tabular}{lcc}
Property & \multicolumn{2}{c}{Lobe} \\
 & West & East \\
\hline
1.5~GHz Flux density (Jy) & 8.7 & 8.1 \\
Volume (cm$^3$) & 6.3$\times$10$^{68}$  & 6.0$\times$10$^{68}$ \\
$B_{min}$ ($\mu$G) & 6.2 & 6.2 \\
P$_{min.en.}$ (\ergpcmcu) & 10$^{-12}$ & 10$^{-12}$ \\
\end{tabular}
\end{table}

Comparing with the IGM pressure at the lobe midpoint
($\sim5\times10^{-12}$\ergpcmcu) the difference in IGM and lobe pressures
of a factor 5 agrees closely with the estimate of \citet{Crostonetal08}.
Under the assumption that P$_{min.en.}$ is an underestimate and the lobes
and IGM are in fact in pressure equilibrium, we can rearrange
Equation~\ref{eqn:psync} to determine the changes in filling factor or
non-radiating particle component necessary to explain the pressure
imbalance. However, we must account for the dependence of $B_{min}$ on k
and $\phi$, and the dependence of $E_e$ on $B_{min}$. In simplified form,
$B_{min} \propto [(1+{\rm k})/\phi]^{1/(\alpha+3)}$, and $E_e \propto
B_{min}^{-(\alpha+1)}$ \citep{WorrallBirkinshaw06}. We find that pressure
balance is achieved if (1+k)/$\phi$=5.1. If, as seems likely, the lobe is
still mildly overpressured with respect to the IGM, this value can be
considered to be a lower limit.

A reduced filling factor could occur in a number of ways; the lobes may
contain a radio-quiet plasma component such as gas entrained by the radio
jets, there may be variation in the relative density of magnetic field and
relativistic particles leading to uneven radio emission, or the lobe volume
could simply be overestimated, for example if their outer surfaces are more
structured than a smooth ellipsoid and the lobes contain pockets or
filaments of IGM plasma.  In principle, it is possible to place a limit on
the effects of the last of these possibilities from the X--ray data. If a
fraction of the apparent lobe volume contains IGM gas, the expected X--ray
surface brightness decrement will be reduced. Unfortunately the IGM surface
brightness is low, only small decrements are expected, and the large
uncertainties mean that we cannot determine a useful limit in this way.
However, the presence of the cavity rims, probably consisting of material
swept up by the expanding lobes, argues that the lobes have been effective
in removing IGM gas from the volume they occupy. A detection of inverse
Compton emission from the lobes (arising from CMB photons scattered up to
X--ray energies by relativistic electrons) might allow us to place limits
on the clumping of the relativistic plasma. Inverse Compton emission has
been detected from the lobes of many FR\,II radio galaxies
\citep{Crostonetal05b} and from the diffuse mini-halo in the core of the
Perseus cluster \citep{SandersFabian07}, but is not detected in the lobes
of bright, cluster-central FR\,I radio sources such as Hydra~A, Hercules~A
and M~87 \citep{HardcastleCroston10, Simionescuetal08}.  Unfortunately, the
inverse Compton emission from the lobes of NGC~4261 is expected to have
only a small flux if the minimum energy condition is correct, $\sim$2~nJy
at 1~keV for each lobe.  This compares with 2.6~nJy measured for the
resolved X--ray jet \citep{Worralletal10}, but the expected 2~nJy would be
spread over the whole area of the lobe and would be undetectable in our
data. We are therefore unable to place limits on the filling factor, and
must continue to assume $\phi$=1.

The additional pressure required could be provided by additional
non--radiating particles in the relativistic plasma. Since we require
k$\sim$4, these would probably have to be introduced into the relativistic
plasma via the entrainment and heating of thermal plasma by the radio jets.
It has been suggested that some knots in the inner jets of Centaurus A may
be caused by interactions between jet plasma and the winds of high mass
stars \citep{Hardcastleetal03}, and stellar mass loss from stars within the
jets is an obvious source of entrained material. The well defined
small--scale X--ray jets have a width of 4.7\arcs\ (720~pc), and length
31.7\arcs\ (4900~pc) for the longer western jet \citep{Worralletal10}, and
we assume they are cylindrical. Using the mean stellar luminosity density
within this radius ($\sim$0.11 \LBsol~pc$^{-3}$, based on optical modelling
described in Section~\ref{sec:CC}) and adopting the stellar mass loss rate
from AGB stars $\dot{M}$=0.0788(\LB/10$^{10}$\LBsol)\Msol~yr$^{-1}$
\citep{Atheyetal02}, we estimate the rate of mass loss into each jet to be
1.67$\times10^{-3}$\Msol~yr$^{-1}$. Assuming the timescale of the current
AGN outburst to be our estimated upper limit (75~Myr), we would thus expect
a minimum of $\sim$125000\Msol\ of stellar material to be entrained by each
jet.

If this entrained material takes the form of a thermal plasma mixed with
the relativistic plasma of the radio lobes, the temperature required for it
to balance the pressure of the surrounding IGM is $>$10~MeV. We note that
\citet{SandersFabian07} place an upper temperature limit of 100~keV on any
thermal plasma in the radio lobes of Perseus~A. For electrons, a
temperature of 10~MeV is equivalent to a Lorentz factor $\gamma\sim20$. The
very low density of such a plasma component ($n_e\sim3\times10^{-7}$\pcmcu)
means that it is unlikely to produce any detectable emission either
directly or via inverse-Compton scattering.  Electrons of $\gamma\sim20$ in
the minimum energy magnetic field would be expected to radiate at
$\sim$1~kHz, below observable frequency bands.

Limits can also be placed on the density of any thermal plasma component in
the lobes, based on the degree of depolarization observed. The relationship
between the observed degree of polarization $\mathcal{P}_{Obs.}$ the the
intrinsic degree of polarization $\mathcal{P}_{Int}$ can be approximated as

\begin{equation}
\noindent \mathcal{P}_{Obs.}=\mathcal{P}_{Int}\frac{\sin(RM\lambda^2)}{RM\lambda^2},
\end{equation}

\noindent where $RM$ is the rotation measure and $\lambda$ the wavelength
\citep{GovoniFeretti04}. The rotation measure of a thermal plasma
intermingled with the relativistic plasma in the radio lobe is
approximately

\begin{equation}
\noindent RM=4.05\times10^5\int\limits_{0}^{L}n_e\, B_\parallel\, dz \,\,\,\,\,  \mathrm{rad\, m^{-2}},
\end{equation}

\noindent where $n_e$ is the electron number density of the thermal
material in cm$^{-3}$, $B_\parallel$ is the line--of--sight component of
the magnetic field in Gauss, and the thickness of the region, $L$, is in pc
\citep{WorrallBirkinshaw06}. A depolarization of $\sim$8 per cent. at
21.2~cm is reported for the lobes of NGC~4261 \citep{Bolognaetal69} and
while the contribution from beam depolarization is unknown, we can limit
the density of any thermal component to be $\la$2$\times$10$^{-4}$\pcmcu.
This is considerably greater than the density expected from entrained gas
from stellar winds (3.6$\times$10$^{-7}$\pcmcu\ for the east lobe) but
comparable to the IGM density. The limit thus argues only against the
entrainment or envelopment of very large quantities of IGM gas into the
radio lobes, but cannot rule out entrainment of gas from stellar winds.

The estimate of the mass of gas from stellar winds entrained by the jets
ignores stars in the outskirts of the galaxy where the jets are broader,
and the possibility of entrainment from the gas of the cool core or IGM.
Increasing the mass entrained would lower the temperature of any thermal
component within the lobes, but the energy required to heat such a
component, $\sim$5$\times10^{57}$~erg summed over both lobes, would not be
significantly altered. This is only $\sim$20 per cent. of the estimated
total mechanical power output of the jets, probably within the
uncertainties on the available energy. Given the evidence of rims around
the cavities, indicating that the lobes probably have a high filling factor
and may be mildly overpressured, it seems likely that the true mechanical
jet power is somewhat higher than our estimate, and this could easily
provide the energy required to heat any entrained material.

\section{Discussion}
\label{sec:discuss}

\subsection{The cool core and AGN}
We can consider three sources of material which might fuel the AGN in
NGC~4261: 1) a cold gas reservoir unrelated to the hot IGM, 2) hot gas
produced within a cool corona, with no significant inflow of gas from the
IGM to the corona, and 3) hot IGM gas via a cooling flow, with no
separation between the cool core and the rest of the IGM. For AGN feedback
to control cooling in the group, the AGN must be fuelled by material which
it is capable of effectively heating, so that its power output directly
affects the rate of fuel supply. This appears unlikely in case 1 and most
likely in case 3, since we observe the interaction between the AGN and IGM.
However, if the cool core is a galactic corona, magnetically separated from
the surrounding IGM, then significant inflow from the IGM cannot occur and
case 3 is ruled out. A feedback loop is possible in case 2 only if the AGN
can effectively heat the gas of the corona or prevent its transport in to
the central engine.

The evidence of past tidal interaction in NGC~4261 raises the question of
whether the current nuclear activity could be related to a merger. This
appears unlikely. Although a nuclear disk of gas and dust containing
(5.4$\pm$1.8)$\times$10$^{4}$\Msol\ of molecular and atomic hydrogen is
observed \citep{Ferrareseetal96}, cool gas has not been detected elsewhere
in NGC~4261 \citep{SerraOosterloo10,Combesetal07}. Stellar population
modelling finds little evidence for a young stellar population in the
galaxy core; the luminosity weighted age is estimated to be
12.6$^{+0.0}_{-0.6}$~Gyr \citep{SerraOosterloo10}. It therefore seems
likely that any merger was gas poor.

While a large injection of cool gas into NGC~4261 is unlikely, it seems
probable that the cool gas disk is the immediate source of material
fuelling the AGN. Direct accretion from the hot phase cannot be ruled out,
but does not explain the presence of the cool gas. Assuming a conversion
efficiency of 1 per cent., accretion of the material in the observed gas
disk would release $\sim$10$^{57}$~erg, a factor of 20 less than the
enthalpy of the cavities. Unless we are coincidentally observing NGC~4261
just as the AGN is about to consume the last of its fuel, the disk gas must
be replenished by material cooling from the hot phase. Judging the
likelihood of such a coincidence is difficult. The selection of NGC~4261 as
a luminous radio galaxy with extended lobes could bias us toward observing
the system during the period when the lobes have had time to expand and
attain a high surface brightness, but before the jets shut down.
Conversely, the lack of evidence of multiple AGN outbursts or star
formation would require any gas introduced by a merger to be efficiently
transported into the galaxy nucleus as a single unit.  However, even if the
current activity were fuelled by cold gas, the presence of the cool core
strongly suggests that the AGN is ineffective in heating its immediate
neighbourhood. The presence of the cool core argues that interaction
between the jets and gas in the core regions is minimal, and shock heating
is weak. The total energy released through shock heating need not be
negligible, but it seems more likely that in this system any recent shocks
have been driven by lobe/jet expansion rather than originating at the
nucleus, and have occurred at radii $>$10~kpc.

The energetic conditions of the cool core (see Section~\ref{sec:CC}) are
similar to those described by \citet{Sunetal07} for galactic coronae.
\citet{Sun09} showed that for systems hosting relatively powerful radio
galaxies, coronae fall into a region of \Lx/$L_{\rm radio}$ space separate
from systems with larger cool cores (see their Figure~1).  From the
deprojection analysis, we estimate the 0.5-2~keV luminosity of the gas in
the core (radius $<$9.8~kpc) to be 5.9$\times$10$^{40}$\ergps. The 1.4~GHz
radio luminosity of 3C~270 is 2.4$\times$10$^{24}$~W~Hz$^{-1}$
\citep{Condonetal02}. This places NGC~4261 in the corona class. However,
the cool core of NGC~4261 is relatively large compared to coronae in galaxy
clusters and has a rather mild temperature difference with the surrounding
IGM. This is expected, as corona size is dependent on the pressure of the
surrounding IGM, which is lower in this group than in higher mass galaxy
clusters.  The limits on the suppression of conduction and efficiency of
AGN heating are also comparatively weak. We conclude that the cool core of
NGC~4261 is probably an example of a nearby, relatively large galactic
corona, but that we cannot be sure of the classification in such a system.

The radio source has clearly affected the structure of the corona.
\citet{Worralletal10} point out that the X--ray jets are surrounded by
regions of low surface brightness, suggesting that thermal gas has been
removed and the lobes now extend back to fill these regions, surrounding
the jets. The X--structure observed in X--ray surface brightness and
temperature maps, and the cool temperature of the structure, suggests that
the gas of the cool core has been compressed and driven away from the jets
by subsonic expansion of the lobes. Direct interaction with the jets
appears unlikely, since there is no evidence of shock heating. The
expansion of the lobes is also probably responsible for the general
north-south extension of the X--ray emission within the stellar body of
NGC~4261; the lobes have pushed the hot IGM gas originally occupying the
outer part of the galaxy away from the jet axis to form an annular
structure which we are viewing edge-on. It is worth noting that such an
annulus would have a similar alignment to the disk of cool gas and dust
which surrounds the AGN, though its scale would be $\sim$200 times larger.

Since the corona shows no signs of having been heated by the current AGN
outburst, it is unclear how nuclear activity can stop cooling in this
region. If the AGN cannot effectively heat the core, it cannot form a
feedback system with the IGM, and cooling and heating of the group gas need
not be in balance. If the feedback relationship is to be sustained, we
require a mechanism for ending nuclear activity. Several possibilities can
be suggested:

\begin{enumerate}
\item If the AGN is fuelled by a reservoir of cool gas left over from a
  gas--rich merger, activity will cease within the next few Myr, assuming a
  constant rate of consumption. This requires us to be observing NGC~4261
  at an unusual period of its history, and suggests that FR\,I radio
  galaxies are fuelled from reservoirs of cool gas, and is thus
  unsatisfying.
\item A change in accretion or jet properties could alter the heating
  mechanism. For example, we could be observing NGC~4261 during a long
  period in which accretion drives jets without producing shocks
  in the nuclear region. At some future point, alteration of conditions
  could cause the amount of energy released in shocks to rise, heating the
  core and ending the outburst. This cannot be ruled out, but we note that
  there is evidence that the $\leq$75~Myr period of activity which has formed the
  radio lobes has not been continuous, and that the jet properties have
  changed with time \citep{Worralletal10}. 
\item The expansion of the lobes back along the line of the jets into the
  corona raises the possibility that they could reach the centre of
  NGC~4261 and envelop the nucleus. This could potentially reduce the
  amount of thermal plasma able to reach the central engine. Enclosure need
  not lead to an immediate cessation of nuclear activity, since the cool
  gas disk would still need to be consumed, and the enclosed region could
  also contain a high--pressure core of hot gas. However, it is unclear
  whether the pressure in the lobes will be sufficient to push aside the
  thermal gas at very small radii. As the lobes expand outward, moving down
  the IGM pressure gradient, we might expect their internal pressure to
  decline. In this case, they are unlikely to be able to move aside the
  high--pressure material in the inner part of the corona. It is also
  unknown whether flows of cooling material might be able to penetrate or
  push aside the enclosure formed by the radio lobes, transporting gas in
  to the nucleus.
\item Disturbance of the structure of the corona by the radio source could
  affect the AGN in a number of ways. The lobes have already compressed the
  gas in the core, increasing its density and cooling rate. This could lead
  to an increase in the supply of gas to the AGN and potentially an
  increased heating rate. Disturbance of the corona structure may also
  affect the conduction rate. The surface of the corona has already been
  increased above that of a sphere. If enough of this surface is in contact
  with the IGM (rather than the radio lobes) the increased surface area
  will lead to an increase in the conduction rate.  Gas motions introduced
  as the lobes push the core gas aside may also alter the magnetic field
  structure, potentially increasing conduction if field lines are
  straightened by expansion, or reducing conduction if the motions cause
  further tangling of the field.  If the suppression factor is
  significantly reduced, the corona could potentially be heated fairly
  rapidly, reducing the fuel supply to the AGN.
\end{enumerate}

While we cannot determine whether any of these processes (or some other)
will be effective in terminating the AGN outburst in NGC~4261, it is likely
that physical changes in the structure of the core over a long timescale
are an important factor. If this is true of other corona systems, it would
suggest that whereas AGN in large cool cores may rapidly reheat the cooling
gas which fuels them, AGN in coronae can remain active for longer periods
without disrupting their fuel supply. Coronae therefore seem likely to fuel
long-term nuclear activity, rather than short outbursts, and AGN in coronae
may spend a larger fraction of their time active than AGN in large cool
cores. This could explain why such systems lack large--scale cool cores;
they inject more energy into the IGM, and heat it to a point where the
cooling which occurs while the AGN is quiescent is insufficient for a cool
core to develop before the jets restart. Studies of other nearby corona
systems are clearly required to determine whether such important
differences in duty cycle and energy output are real. \citet{Dunnetal10}
examined 18 nearby giant elliptical galaxies and found that 17 host nuclear
radio sources, potentially lending weight to the idea that AGN in galactic
coronae are able to remain active for long periods. However, a number of
galaxies in the sample occupy the centres of groups or clusters, and have
cool cores considerably larger than that of NGC~4261, so only a fraction of
this sample are likely to be corona systems.

\subsection{Structure of the radio source}
The structure of the radio lobes, cavities and rims raises several
questions. Considerably more X--ray emission is seen in the eastern lobe
and its rim appears, at least in the \xmm\ images, to be thicker. It is
also complete, unlike the western rim, which is undetected in one quadrant.
Two possible explanations involve the data quality; the \xmm\ data may
simply be insufficiently deep to clearly detect the cavity rims and
determine their shape, and our point source subtraction may also be
ineffective. Prior studies of the X--ray point source population have shown
a particular concentration of sources to the north and northeast of the AGN
\citep{Giordanoetal05}, but also to the south and west. The poorer ability
of \xmm\ to resolve these sources compared to \chandra\ might produce
increased noise in the northeast quadrant, leading us to overestimate the
thickness of the rim in this area.

Other factors which could affect rim strength are the structure of the IGM
into which the radio lobes have expanded, and the structure of the lobes
themselves. Surface brightness modelling of the group with the \rosat\ PSPC
found the IGM to be centred to the east of NGC~4261, with the emission more
extended to the south and east. \citet{Davisetal95} found the centroid of
diffuse emission to be R.A. 12$^h$19$^m$27$^s$, Dec.
+05\degree49\arcm58\arcs\ (J2000), $\sim$1\arcm\ (9.2~kpc) ENE of the
nucleus (see also their fig.~1). If this extension is real, it suggests
that NGC~4261 is slightly offset from the centre of the group X--ray halo.
The eastern lobe would, in this scenario, have expanded into a denser
environment than its western counterpart, and has therefore accumulated
more gas in its rim.

The morphologies of the two radio lobes are also slightly different. The
western lobe is broader but less extended, and has distinct bulges on its
sides.  If the extension of the lobe tip is driven by the jet, the shorter
length of the western lobe could indicate that the western jet has
encountered greater resistance to its expansion. However, we see no
evidence of any feature which could be responsible for this resistance,
such as a region of denser intra--group gas. The greater sideways expansion
of the western lobe suggests that external pressure is lower on this side
of the core, as expected if the galaxy is offset from the group centre.
There is also some indication that the eastern radio jet bends inside the
radio lobe \citep[see the contour map in Figure~\ref{fig:4ims}a and fig.~1
of][]{Worralletal10}, while the western jet remains on approximately the
same axis as the small-scale jets visible in the X--ray. This could perhaps
affect the apparent structure of the eastern lobe, since we are viewing a
projected image of a three dimensional structure. If the jet axis is no
longer in the plane of the sky, the shell of gas around the expanding lobe
tip might appear as a broader X--ray bright region rather than a relatively
thin shell. This would probably require a fairly strong bend in the jet
which, while quite possible, has yet to be confirmed from the available
observational data.

\section{Conclusions}
\label{sec:conc}
We have analysed deep \chandra\ and \xmm\ observations of the nearby
group--central elliptical NGC~4261. The X--ray observations reveal a large
degree of structure in the gas in which the galaxy is embedded. The AGN
jets have inflated two large lobes, excavating cavities in the IGM and
building up rims of compressed hot gas which almost entirely enclose the
lobes. The western cavity has an apparent break in its rim, while the
eastern lobe has a higher X--ray luminosity over most of its area. The
enthalpy of the lobes is large, $\sim$2.4$\times$10$^{58}$~erg, and the
likely timescale of the outburst, $\leq$75~Myr, suggests that the
mechanical power of the jets, $\geq$10$^{43}$\ergps, greatly exceeds their
current radio power. The mechanical power of the AGN also exceeds the
X--ray luminosity of the gas in the region the lobes inhabit, the central
45~kpc of the group.  The available X--ray data are insufficient to
determine conclusively whether the cavity rims contain shocks, but we can
limit the maximum expansion velocity of the lobes to be
$\mathcal{M}\leq1.05$, and their minimum expansion velocity to be a large
fraction of the sound speed.  These results suggest that the radio source
may still be mildly overpressured relative to it environment, but that it
is only inefficiently heating the IGM.

Spectral mapping and radial spectral analysis confirms that NGC~4261 has a
cool core with radius $\sim$9.8~kpc and a typical temperature of
$\sim$0.6~keV, compared to a peak IGM temperature of $\sim$1.6~keV. The
core is not spherically symmetric, but contains an X--shaped structure,
with the arms of the X extending to either side of the radio jets, the
northeast and southwest arms being most extended. This structure was
commented on by \citet{Worralletal10}, who noted that the radio jets appear
to have expelled gas in wedge--shaped regions to their north and south,
probably corresponding to conical volumes. There is also a larger,
bar--shaped structure extending north--south across the centre of NGC~4261,
which is poorly correlated with the galaxy major axis but perpendicular to
the jet axis. This is probably an annulus of gas viewed edge on, pushed
into its current configuration by the expansion of the radio lobes.  It is
clear from these structures that the jets have played an important role in
determining the structure and properties of the gas in and around NGC~4261.

The properties of the cool core suggest that it should be classified as a
galactic corona. The relative radio power of the galaxy and luminosity of
the cool core fall in the range occupied by corona systems. Heating of the
corona by the AGN and conduction must have relatively low efficiencies. We
limit the AGN heating efficiency to be $<$3.5 per cent. and find that
conduction into the cool core from the IGM is suppressed by a factor $>$5.
However these values probably overestimate the effectiveness of the heating
mechanisms since they assume a timescale based on the length of the AGN
outburst rather than the lifetime of the temperature gradient. Supernovae
within the corona release $\sim$6$\times10^{40}$\ergps (compared to the
0.3-7.0~keV X--ray gas luminosity of $\sim$10$^{41}$\ergps), but since the
stellar distribution is less centrally peaked than the gas density, they
are unlikely to be able to prevent cooling in the inner core.  Material
lost from stars can balance the cooling rate of the corona, suggesting that
cooling could fuel the AGN over timescales of several Gyr.

Since the AGN appears unable to effectively heat the corona, we have
considered alternative mechanisms by which cooling could be disrupted or
stopped. We conclude that changes in the corona structure, caused by the
intrusion of the radio lobes into the cooling region, may be able to alter
the cooling rate by compressing the gas, isolating the nucleus, or
increasing the conduction rate. In any case, these results suggest that
corona systems can potentially sustain longer periods of nuclear activity
compared to their counterparts in large cool cores. Since the powers of
radio galaxies with coronae and large cool cores are similar, it seems
likely that sources in coronae will, on average, release more energy into
the IGM over long timescales. Given the large numbers of coronae already
detected in group-- and cluster--central galaxies \citep[e.g.,][]{Sun09},
this may be an important consideration in determining the gas temperature
structure of the population of galaxy systems.

Comparing the pressures of the IGM gas and the relativistic plasma of the
lobes, we find the apparent lobe pressure to be a factor of 5 lower than
that of the surrounding halo. This indicates that either additional
non--radiating particles are present in the lobe plasma, that the filling
factor of the lobes is low, or that the magnetic field departs from the
minimum--energy solution. The currently available data are insufficient to
place useful limits on the filling factor of the lobes. Thermal gas
entrained from stars within the radio jets, or from the IGM, could
potentially provide the additional pressure without radiating in a
detectable waveband or producing excessive polarization effects.  The
energy required to heat this gas to its presumed temperature is $\sim$20
per cent. of the total enthalpy of the radio lobes.

\medskip
\noindent\textbf{Acknowledgements}\\
The authors thank the anonymous referee for several useful suggestions, and
P. Bonfini for providing optical data for NGC~4261. EO'S thanks A.
Sanderson for useful discussions of the X-ray analysis.  Support for this
work was provided by the National Aeronautics and Space Administration
through Chandra Award Number G08-9094X issued by the Chandra X-ray
Observatory Center, which is operated by the Smithsonian Astrophysical
Observatory for and on behalf of the National Aeronautics Space
Administration under contract NAS8-03060. This work is partially based on
observations obtained with XMM-Newton, an ESA science mission with
instruments and contributions directly funded by ESA Member States and
NASA. This work has used data from the VLA. EO'S acknowledges the support
of the European Community under the Marie Curie Research Training Network.
Space Astrophysics in Crete is partly supported by EU FP7
\textit{Capacities} GA No206469. GT and AW acknowledge support from the
Agenzia Spaziale Italiana through grant ASI-INAF I/009/10/0. We acknowledge
the usage of the HyperLeda database (http://leda.univ-lyon1.fr).


\bibliographystyle{mn2e}
\bibliography{../paper}

\begin{thebibliography}{}

\bibitem[\protect\citeauthoryear{{Allen}, {Dunn}, {Fabian}, {Taylor} \&
  {Reynolds}}{{Allen} et~al.}{2006}]{Allenetal06}
{Allen} S.~W.,  {Dunn} R.~J.~H.,  {Fabian} A.~C.,  {Taylor} G.~B.,
  {Reynolds} C.~S.,  2006, MNRAS, 372, 21

\bibitem[\protect\citeauthoryear{{Arnaud}, {Majerowicz}, {Lumb}, {Neumann},
  {Aghanim}, {Blanchard}, {Boer}, {Burke}, {Collins}, {Giard}, {Nevalainen},
  {Nichol}, {Romer} \& {Sadat}}{{Arnaud} et~al.}{2002}]{Arnaudetal02}
{Arnaud} M.,  {Majerowicz} S.,  {Lumb} D.,  {Neumann} D.~M.,  {Aghanim} N.,
  {Blanchard} A.,  {Boer} M.,  {Burke} D.~J.,  {Collins} C.~A.,  {Giard} M.,
  {Nevalainen} J.,  {Nichol} R.~C.,  {Romer} A.~K.,    {Sadat} R.,  2002, A\&A,
  390, 27

\bibitem[\protect\citeauthoryear{{Athey}, {Bregman}, {Bregman}, {Temi} \&
  {Sauvage}}{{Athey} et~al.}{2002}]{Atheyetal02}
{Athey} A.,  {Bregman} J.,  {Bregman} J.,  {Temi} P.,    {Sauvage} M.,  2002,
  ApJ, 571, 272

\bibitem[\protect\citeauthoryear{{Balmaverde}, {Baldi} \&
  {Capetti}}{{Balmaverde} et~al.}{2008}]{Balmaverdeetal08}
{Balmaverde} B.,  {Baldi} R.~D.,    {Capetti} A.,  2008, A\&A, 486, 119

\bibitem[\protect\citeauthoryear{{Balmaverde}, {Capetti} \&
  {Grandi}}{{Balmaverde} et~al.}{2006}]{Balmaverdeetal06}
{Balmaverde} B.,  {Capetti} A.,    {Grandi} P.,  2006, A\&A, 451, 35

\bibitem[\protect\citeauthoryear{{B{\^i}rzan}, {McNamara}, {Nulsen}, {Carilli}
  \& {Wise}}{{B{\^i}rzan} et~al.}{2008}]{Birzanetal08}
{B{\^i}rzan} L.,  {McNamara} B.~R.,  {Nulsen} P.~E.~J.,  {Carilli} C.~L.,
  {Wise} M.~W.,  2008, ApJ, 686, 859

\bibitem[\protect\citeauthoryear{{Blanton}, {Randall}, {Douglass}, {Sarazin},
  {Clarke} \& {McNamara}}{{Blanton} et~al.}{2009}]{Blantonetal09}
{Blanton} E.~L.,  {Randall} S.~W.,  {Douglass} E.~M.,  {Sarazin} C.~L.,
  {Clarke} T.~E.,    {McNamara} B.~R.,  2009, ApJ, 697, L95

\bibitem[\protect\citeauthoryear{{B{\"o}hringer}, {Briel}, {Schwarz}, {Voges},
  {Hartner} \& {Tr{\"u}mper}}{{B{\"o}hringer} et~al.}{1994}]{Bohringeretal94}
{B{\"o}hringer} H.,  {Briel} U.~G.,  {Schwarz} R.~A.,  {Voges} W.,  {Hartner}
  G.,    {Tr{\"u}mper} J.,  1994, Nature, 368, 828

\bibitem[\protect\citeauthoryear{{Bologna}, {McClain} \& {Sloanaker}}{{Bologna}
  et~al.}{1969}]{Bolognaetal69}
{Bologna} J.~M.,  {McClain} E.~F.,    {Sloanaker} R.~M.,  1969, ApJ, 156, 815

\bibitem[\protect\citeauthoryear{{Bonfini}, {Zezas} \& {et al.}}{{Bonfini}
  et~al.}{2011}]{Bonfinietal11}
{Bonfini} P.,  {Zezas} A.,    {et al.} 2011, in prep.

\bibitem[\protect\citeauthoryear{{Burns}}{{Burns}}{1990}]{Burns90}
{Burns} J.~O.,  1990, AJ, 99, 14

\bibitem[\protect\citeauthoryear{{Burns}, {Owen} \& {Rudnick}}{{Burns}
  et~al.}{1979}]{Burnsetal79}
{Burns} J.~O.,  {Owen} F.~N.,    {Rudnick} L.,  1979, AJ, 84, 1683

\bibitem[\protect\citeauthoryear{{Cappellari}, {Emsellem}, {Bacon}, {Bureau},
  {Davies}, {de Zeeuw}, {Falcon-Barroso}, {Krajnovic}, {Kuntschner},
  {McDermid}, {Peletier}, {Sarzi}, {van den Bosch} \& {van de
  Ven}}{{Cappellari} et~al.}{2007}]{Cappellarietal07}
{Cappellari} M.,  {Emsellem} E.,  {Bacon} R.,  {Bureau} M.,  {Davies} R.~L.,
  {de Zeeuw} P.~T.,  {Falcon-Barroso} J.,  {Krajnovic} D.,  {Kuntschner} H.,
  {McDermid} R.~M.,  {Peletier} R.~F.,  {Sarzi} M.,  {van den Bosch} R.~C.~E.,
    {van de Ven} G.,  2007, MNRAS, 379, 418

\bibitem[\protect\citeauthoryear{Cappellaro, Evans \& Turatto}{Cappellaro
  et~al.}{1999}]{Cappellaroetal99}
Cappellaro E.,  Evans R.,    Turatto M.,  1999, A\&A, 351, 459

\bibitem[\protect\citeauthoryear{{Cavagnolo}, {McNamara}, {Nulsen}, {Carilli},
  {Jones} \& {Birzan}}{{Cavagnolo} et~al.}{2010}]{Cavagnoloetal10}
{Cavagnolo} K.~W.,  {McNamara} B.~R.,  {Nulsen} P.~E.~J.,  {Carilli} C.~L.,
  {Jones} C.,    {Birzan} L.,  2010, ApJ, 720, 1066

\bibitem[\protect\citeauthoryear{{Chakrabarty} \& {Jackson}}{{Chakrabarty} \&
  {Jackson}}{2009}]{ChakrabartyJackson09}
{Chakrabarty} D.,  {Jackson} B.,  2009, A\&A, 498, 615

\bibitem[\protect\citeauthoryear{{Chiaberge}, {Capetti} \&
  {Celotti}}{{Chiaberge} et~al.}{1999}]{Chiabergeetal99}
{Chiaberge} M.,  {Capetti} A.,    {Celotti} A.,  1999, A\&A, 349, 77

\bibitem[\protect\citeauthoryear{{Churazov}, {Br{\" u}ggen}, {Kaiser}, {B{\"
  o}hringer} \& {Forman}}{{Churazov} et~al.}{2001}]{Churazovetal01}
{Churazov} E.,  {Br{\" u}ggen} M.,  {Kaiser} C.~R.,  {B{\" o}hringer} H.,
  {Forman} W.,  2001, ApJ, 554, 261

\bibitem[\protect\citeauthoryear{{Combes}, {Young} \& {Bureau}}{{Combes}
  et~al.}{2007}]{Combesetal07}
{Combes} F.,  {Young} L.~M.,    {Bureau} M.,  2007, MNRAS, 377, 1795

\bibitem[\protect\citeauthoryear{{Condon}, {Cotton} \& {Broderick}}{{Condon}
  et~al.}{2002}]{Condonetal02}
{Condon} J.~J.,  {Cotton} W.~D.,    {Broderick} J.~J.,  2002, AJ, 124, 675

\bibitem[\protect\citeauthoryear{{Cowie} \& {McKee}}{{Cowie} \&
  {McKee}}{1977}]{CowieMcKee77}
{Cowie} L.~L.,  {McKee} C.~F.,  1977, ApJ, 211, 135

\bibitem[\protect\citeauthoryear{{Croston}, {Hardcastle} \&
  {Birkinshaw}}{{Croston} et~al.}{2005}]{Crostonetal05}
{Croston} J.~H.,  {Hardcastle} M.~J.,    {Birkinshaw} M.,  2005, MNRAS, 357,
  279

\bibitem[\protect\citeauthoryear{{Croston}, {Hardcastle}, {Birkinshaw},
  {Worrall} \& {Laing}}{{Croston} et~al.}{2008}]{Crostonetal08}
{Croston} J.~H.,  {Hardcastle} M.~J.,  {Birkinshaw} M.,  {Worrall} D.~M.,
  {Laing} R.~A.,  2008, MNRAS, 386, 1709

\bibitem[\protect\citeauthoryear{{Croston}, {Hardcastle}, {Harris}, {Belsole},
  {Birkinshaw} \& {Worrall}}{{Croston} et~al.}{2005}]{Crostonetal05b}
{Croston} J.~H.,  {Hardcastle} M.~J.,  {Harris} D.~E.,  {Belsole} E.,
  {Birkinshaw} M.,    {Worrall} D.~M.,  2005, ApJ, 626, 733

\bibitem[\protect\citeauthoryear{{Croston}, {Kraft}, {Hardcastle},
  {Birkinshaw}, {Worrall}, {Nulsen}, {Penna}, {Sivakoff}, {Jord{\'a}n},
  {Brassington}, {Evans}, {Forman}, {Gilfanov}, {Goodger}, {Harris} \& et
  al.}{{Croston} et~al.}{2009}]{Crostonetal09}
{Croston} J.~H.,  {Kraft} R.~P.,  {Hardcastle} M.~J.,  {Birkinshaw} M.,
  {Worrall} D.~M.,  {Nulsen} P.~E.~J.,  {Penna} R.~F.,  {Sivakoff} G.~R.,
  {Jord{\'a}n} A.,  {Brassington} N.~J.,  {Evans} D.~A.,  {Forman} W.~R.,
  {Gilfanov} M.,  {Goodger} J.~L.,  {Harris} W.~E.,    et al. 2009, MNRAS, 395,
  1999

\bibitem[\protect\citeauthoryear{{David}, {Jones}, {Forman}, {Nulsen},
  {Vrtilek}, {O'Sullivan}, {Giacintucci} \& {Raychaudhury}}{{David}
  et~al.}{2009}]{Davidetal09}
{David} L.~P.,  {Jones} C.,  {Forman} W.,  {Nulsen} P.,  {Vrtilek} J.,
  {O'Sullivan} E.,  {Giacintucci} S.,    {Raychaudhury} S.,  2009, ApJ, 705,
  624

\bibitem[\protect\citeauthoryear{{Davies} \& {Birkinshaw}}{{Davies} \&
  {Birkinshaw}}{1986}]{DaviesBirkinshaw86}
{Davies} R.~L.,  {Birkinshaw} M.,  1986, ApJ, 303, L45

\bibitem[\protect\citeauthoryear{{Davis}, {Mushotzky}, {Mulchaey}, {Worrall},
  {Birkinshaw} \& {Burstein}}{{Davis} et~al.}{1995}]{Davisetal95}
{Davis} D.~S.,  {Mushotzky} R.~F.,  {Mulchaey} J.~S.,  {Worrall} D.~M.,
  {Birkinshaw} M.,    {Burstein} D.,  1995, ApJ, 444, 582

\bibitem[\protect\citeauthoryear{{Dickey} \& {Lockman}}{{Dickey} \&
  {Lockman}}{1990}]{DickeyLockman90}
{Dickey} J.~M.,  {Lockman} F.~J.,  1990, ARA\&A, 28, 215

\bibitem[\protect\citeauthoryear{{Dong}, {Rasmussen} \& {Mulchaey}}{{Dong}
  et~al.}{2010}]{Dongetal10}
{Dong} R.,  {Rasmussen} J.,    {Mulchaey} J.~S.,  2010, ApJ, 712, 883

\bibitem[\protect\citeauthoryear{{Dunn}, {Allen}, {Taylor}, {Shurkin},
  {Gentile}, {Fabian} \& {Reynolds}}{{Dunn} et~al.}{2010}]{Dunnetal10}
{Dunn} R.~J.~H.,  {Allen} S.~W.,  {Taylor} G.~B.,  {Shurkin} K.~F.,  {Gentile}
  G.,  {Fabian} A.~C.,    {Reynolds} C.~S.,  2010, MNRAS, 404, 180

\bibitem[\protect\citeauthoryear{{Dunn}, {Fabian} \& {Taylor}}{{Dunn}
  et~al.}{2005}]{Dunnetal05}
{Dunn} R.~J.~H.,  {Fabian} A.~C.,    {Taylor} G.~B.,  2005, MNRAS, 364, 1343

\bibitem[\protect\citeauthoryear{{Ellis} \& {O'Sullivan}}{{Ellis} \&
  {O'Sullivan}}{2006}]{EllisOSullivan06}
{Ellis} S.~C.,  {O'Sullivan} E.,  2006, MNRAS, 367, 627

\bibitem[\protect\citeauthoryear{{Emonts}, {Morganti}, {Oosterloo}, {van der
  Hulst}, {van Moorsel} \& {Tadhunter}}{{Emonts} et~al.}{2007}]{Emontsetal07}
{Emonts} B.~H.~C.,  {Morganti} R.,  {Oosterloo} T.~A.,  {van der Hulst} J.~M.,
  {van Moorsel} G.,    {Tadhunter} C.~N.,  2007, A\&A, 464, L1

\bibitem[\protect\citeauthoryear{{Fabian}, {Allen}, {Crawford}, {Johnstone},
  {Morris}, {Sanders} \& {Schmidt}}{{Fabian} et~al.}{2002}]{Fabianetal02}
{Fabian} A.~C.,  {Allen} S.~W.,  {Crawford} C.~S.,  {Johnstone} R.~M.,
  {Morris} R.~G.,  {Sanders} J.~S.,    {Schmidt} R.~W.,  2002, MNRAS, 332, L50

\bibitem[\protect\citeauthoryear{{Fabian}, {Sanders}, {Taylor} \&
  {Allen}}{{Fabian} et~al.}{2005}]{Fabianetal05}
{Fabian} A.~C.,  {Sanders} J.~S.,  {Taylor} G.~B.,    {Allen} S.~W.,  2005,
  MNRAS, 360, L20

\bibitem[\protect\citeauthoryear{{Fabian}, {Sanders}, {Taylor}, {Allen},
  {Crawford}, {Johnstone} \& {Iwasawa}}{{Fabian} et~al.}{2006}]{Fabianetal06}
{Fabian} A.~C.,  {Sanders} J.~S.,  {Taylor} G.~B.,  {Allen} S.~W.,  {Crawford}
  C.~S.,  {Johnstone} R.~M.,    {Iwasawa} K.,  2006, MNRAS, 366, 417

\bibitem[\protect\citeauthoryear{{Feretti}, {Perola} \& {Fanti}}{{Feretti}
  et~al.}{1992}]{Ferettietal92}
{Feretti} L.,  {Perola} G.~C.,    {Fanti} R.,  1992, A\&A, 265, 9

\bibitem[\protect\citeauthoryear{{Ferrarese}, {Ford} \& {Jaffe}}{{Ferrarese}
  et~al.}{1996}]{Ferrareseetal96}
{Ferrarese} L.,  {Ford} H.~C.,    {Jaffe} W.,  1996, ApJ, 470, 444

\bibitem[\protect\citeauthoryear{{Forman}, {Jones}, {Churazov}, {Markevitch},
  {Nulsen}, {Vikhlinin}, {Begelman}, {B{\"o}hringer}, {Eilek}, {Heinz},
  {Kraft}, {Owen} \& {Pahre}}{{Forman} et~al.}{2007}]{Formanetal07}
{Forman} W.,  {Jones} C.,  {Churazov} E.,  {Markevitch} M.,  {Nulsen} P.,
  {Vikhlinin} A.,  {Begelman} M.,  {B{\"o}hringer} H.,  {Eilek} J.,  {Heinz}
  S.,  {Kraft} R.,  {Owen} F.,    {Pahre} M.,  2007, ApJ, 665, 1057

\bibitem[\protect\citeauthoryear{{Freeman}, {Doe} \& {Siemiginowska}}{{Freeman}
  et~al.}{2001}]{Freemanetal01}
{Freeman} P.,  {Doe} S.,    {Siemiginowska} A.,  2001, in {J.-L.~Starck \&
  F.~D.~Murtagh} ed., Society of Photo-Optical Instrumentation Engineers (SPIE)
  Conference Series Vol.~4477 of Society of Photo-Optical Instrumentation
  Engineers (SPIE) Conference Series, {Sherpa: a mission-independent data
  analysis application}.
p.~76

\bibitem[\protect\citeauthoryear{{Giordano}, {Cortese}, {Trinchieri}, {Wolter},
  {Colpi}, {Gavazzi} \& {Mayer}}{{Giordano} et~al.}{2005}]{Giordanoetal05}
{Giordano} L.,  {Cortese} L.,  {Trinchieri} G.,  {Wolter} A.,  {Colpi} M.,
  {Gavazzi} G.,    {Mayer} L.,  2005, ApJ, 634, 272

\bibitem[\protect\citeauthoryear{{Gliozzi}, {Sambruna} \& {Brandt}}{{Gliozzi}
  et~al.}{2003}]{Gliozzietal03}
{Gliozzi} M.,  {Sambruna} R.~M.,    {Brandt} W.~N.,  2003, A\&A, 408, 949

\bibitem[\protect\citeauthoryear{{Govoni} \& {Feretti}}{{Govoni} \&
  {Feretti}}{2004}]{GovoniFeretti04}
{Govoni} F.,  {Feretti} L.,  2004, International Journal of Modern Physics D,
  13, 1549

\bibitem[\protect\citeauthoryear{{Grevesse} \& {Sauval}}{{Grevesse} \&
  {Sauval}}{1998}]{GrevesseSauval98}
{Grevesse} N.,  {Sauval} A.~J.,  1998, Space Sci.~Rev., 85, 161

\bibitem[\protect\citeauthoryear{{Hardcastle} \& {Croston}}{{Hardcastle} \&
  {Croston}}{2010}]{HardcastleCroston10}
{Hardcastle} M.~J.,  {Croston} J.~H.,  2010, MNRAS, 404, 2018

\bibitem[\protect\citeauthoryear{{Hardcastle}, {Evans} \&
  {Croston}}{{Hardcastle} et~al.}{2007}]{Hardcastleetal07}
{Hardcastle} M.~J.,  {Evans} D.~A.,    {Croston} J.~H.,  2007, MNRAS, 376, 1849

\bibitem[\protect\citeauthoryear{{Hardcastle} \& {Worrall}}{{Hardcastle} \&
  {Worrall}}{2000}]{HardcastleWorrall00}
{Hardcastle} M.~J.,  {Worrall} D.~M.,  2000, MNRAS, 319, 562

\bibitem[\protect\citeauthoryear{{Hardcastle}, {Worrall}, {Kraft}, {Forman},
  {Jones} \& {Murray}}{{Hardcastle} et~al.}{2003}]{Hardcastleetal03}
{Hardcastle} M.~J.,  {Worrall} D.~M.,  {Kraft} R.~P.,  {Forman} W.~R.,  {Jones}
  C.,    {Murray} S.~S.,  2003, ApJ, 593, 169

\bibitem[\protect\citeauthoryear{Helsdon \& Ponman}{Helsdon \&
  Ponman}{2000}]{Helsdonponman00}
Helsdon S.~F.,  Ponman T.~J.,  2000, MNRAS, 315, 356

\bibitem[\protect\citeauthoryear{{Ho}, {Filippenko} \& {Sargent}}{{Ho}
  et~al.}{1997}]{Hoetal97b}
{Ho} L.~C.,  {Filippenko} A.~V.,    {Sargent} W.~L.~W.,  1997, ApJS, 112, 315

\bibitem[\protect\citeauthoryear{{Humphrey}, {Buote}, {Brighenti}, {Gebhardt}
  \& {Mathews}}{{Humphrey} et~al.}{2009}]{Humphreyetal09b}
{Humphrey} P.~J.,  {Buote} D.~A.,  {Brighenti} F.,  {Gebhardt} K.,    {Mathews}
  W.~G.,  2009, ApJ, 703, 1257

\bibitem[\protect\citeauthoryear{{Humphrey}, {Buote}, {Gastaldello},
  {Zappacosta}, {Bullock}, {Brighenti} \& {Mathews}}{{Humphrey}
  et~al.}{2006}]{Humphreyetal06}
{Humphrey} P.~J.,  {Buote} D.~A.,  {Gastaldello} F.,  {Zappacosta} L.,
  {Bullock} J.~S.,  {Brighenti} F.,    {Mathews} W.~G.,  2006, ApJ, 646, 899

\bibitem[\protect\citeauthoryear{{Humphrey}, {Liu} \& {Buote}}{{Humphrey}
  et~al.}{2009}]{Humphreyetal09}
{Humphrey} P.~J.,  {Liu} W.,    {Buote} D.~A.,  2009, ApJ, 693, 822

\bibitem[\protect\citeauthoryear{{Jaffe}, {Ford}, {Ferrarese}, {van den Bosch}
  \& {O'Connell}}{{Jaffe} et~al.}{1993}]{Jaffeetal93}
{Jaffe} W.,  {Ford} H.~C.,  {Ferrarese} L.,  {van den Bosch} F.,    {O'Connell}
  R.~W.,  1993, Nature, 364, 213

\bibitem[\protect\citeauthoryear{{Jaffe} \& {McNamara}}{{Jaffe} \&
  {McNamara}}{1994}]{Jaffeetal94}
{Jaffe} W.,  {McNamara} B.~R.,  1994, ApJ, 434, 110

\bibitem[\protect\citeauthoryear{{Jansen}, {Lumb}, {Altieri}, {Clavel}, {Ehle},
  {Erd}, {Gabriel}, {Guainazzi}, {Gondoin}, {Much}, {Munoz}, {Santos},
  {Schartel}, {Texier} \& {Vacanti}}{{Jansen} et~al.}{2001}]{Jansenetal01}
{Jansen} F.,  {Lumb} D.,  {Altieri} B.,  {Clavel} J.,  {Ehle} M.,  {Erd} C.,
  {Gabriel} C.,  {Guainazzi} M.,  {Gondoin} P.,  {Much} R.,  {Munoz} R.,
  {Santos} M.,  {Schartel} N.,  {Texier} D.,    {Vacanti} G.,  2001, A\&A, 365,
  L1

\bibitem[\protect\citeauthoryear{{Kalberla}, {Burton}, {Hartmann}, {Arnal},
  {Bajaja}, {Morras} \& {P{\"o}ppel}}{{Kalberla} et~al.}{2005}]{Kalberlaetal05}
{Kalberla} P.~M.~W.,  {Burton} W.~B.,  {Hartmann} D.,  {Arnal} E.~M.,  {Bajaja}
  E.,  {Morras} R.,    {P{\"o}ppel} W.~G.~L.,  2005, A\&A, 440, 775

\bibitem[\protect\citeauthoryear{{Kim} \& {Fabbiano}}{{Kim} \&
  {Fabbiano}}{2004}]{KimFabbiano04}
{Kim} D.,  {Fabbiano} G.,  2004, ApJ, 611, 846

\bibitem[\protect\citeauthoryear{{K\"{u}hr}, {Witzel}, {Pauliny-Toth} \&
  {Nauber}}{{K\"{u}hr} et~al.}{1981}]{Kuhretal81}
{K\"{u}hr} H.,  {Witzel} A.,  {Pauliny-Toth} I.~I.~K.,    {Nauber} U.,  1981,
  A\&AS, 45, 367

\bibitem[\protect\citeauthoryear{{Kuntz} \& {Snowden}}{{Kuntz} \&
  {Snowden}}{2000}]{KuntzSnowden00}
{Kuntz} K.~D.,  {Snowden} S.~L.,  2000, ApJ, 543, 195

\bibitem[\protect\citeauthoryear{{Landau} \& {Lifshitz}}{{Landau} \&
  {Lifshitz}}{1959}]{LandauLifshitz59}
{Landau} L.,  {Lifshitz} E.,  1959, Fluid mechanics.
Oxford: Pergamon Press

\bibitem[\protect\citeauthoryear{{Leahy}}{{Leahy}}{1991}]{Leahy91}
{Leahy} J.~P.,  1991, {Interpretation of large scale extragalactic jets, in
  Beams and Jets in Astrophysics, ed. Hughes, P.A.}.
Cambridge University Press, p.~100

\bibitem[\protect\citeauthoryear{{Matsumoto}, {Fukazawa}, {Nakazawa}, {Iyomoto}
  \& {Makishima}}{{Matsumoto} et~al.}{2001}]{Matsumotoetal01}
{Matsumoto} Y.,  {Fukazawa} Y.,  {Nakazawa} K.,  {Iyomoto} N.,    {Makishima}
  K.,  2001, PASJ, 53, 475

\bibitem[\protect\citeauthoryear{{McNamara} \& {Nulsen}}{{McNamara} \&
  {Nulsen}}{2007}]{McNamaraNulsen07}
{McNamara} B.~R.,  {Nulsen} P.~E.~J.,  2007, ARA\&A, 45, 117

\bibitem[\protect\citeauthoryear{{McNamara}, {Rohanizadegan} \&
  {Nulsen}}{{McNamara} et~al.}{2011}]{McNamaraetal11}
{McNamara} B.~R.,  {Rohanizadegan} M.,    {Nulsen} P.~E.~J.,  2011, ApJ, 727,
  39

\bibitem[\protect\citeauthoryear{{Million}, {Werner}, {Simionescu}, {Allen},
  {Nulsen}, {Fabian}, {Bohringer} \& {Sanders}}{{Million}
  et~al.}{2010}]{Millionetal10}
{Million} E.~T.,  {Werner} N.,  {Simionescu} A.,  {Allen} S.~W.,  {Nulsen}
  P.~E.~J.,  {Fabian} A.~C.,  {Bohringer} H.,    {Sanders} J.~S.,  2010, MNRAS,
  407, 2046

\bibitem[\protect\citeauthoryear{{Mittal}, {Hudson}, {Reiprich} \&
  {Clarke}}{{Mittal} et~al.}{2009}]{Mittaletal09}
{Mittal} R.,  {Hudson} D.~S.,  {Reiprich} T.~H.,    {Clarke} T.,  2009, A\&A,
  501, 835

\bibitem[\protect\citeauthoryear{{Nolthenius}}{{Nolthenius}}{1993}]{Nolthenius%
93}
{Nolthenius} R.,  1993, ApJS, 85, 1

\bibitem[\protect\citeauthoryear{{Nulsen}, {McNamara}, {Wise} \&
  {David}}{{Nulsen} et~al.}{2005}]{Nulsenetal05}
{Nulsen} P.~E.~J.,  {McNamara} B.~R.,  {Wise} M.~W.,    {David} L.~P.,  2005,
  ApJ, 628, 629

\bibitem[\protect\citeauthoryear{{O'Sullivan}, {Giacintucci}, {David},
  {Vrtilek} \& {Raychaudhury}}{{O'Sullivan} et~al.}{2010}]{OSullivanetal10}
{O'Sullivan} E.,  {Giacintucci} S.,  {David} L.~P.,  {Vrtilek} J.~M.,
  {Raychaudhury} S.,  2010, MNRAS, 407, 321

\bibitem[\protect\citeauthoryear{{O'Sullivan}, {Giacintucci}, {Vrtilek},
  {Raychaudhury} \& {David}}{{O'Sullivan} et~al.}{2009}]{OSullivanetal09}
{O'Sullivan} E.,  {Giacintucci} S.,  {Vrtilek} J.~M.,  {Raychaudhury} S.,
  {David} L.~P.,  2009, ApJ, 701, 1560

\bibitem[\protect\citeauthoryear{{O'Sullivan}, {Vrtilek}, {Harris} \&
  {Ponman}}{{O'Sullivan} et~al.}{2007}]{OSullivanetal07}
{O'Sullivan} E.,  {Vrtilek} J.~M.,  {Harris} D.~E.,    {Ponman} T.~J.,  2007,
  ApJ, 658, 299

\bibitem[\protect\citeauthoryear{{Parrish} \& {Quataert}}{{Parrish} \&
  {Quataert}}{2008}]{ParrishQuataert08}
{Parrish} I.~J.,  {Quataert} E.,  2008, ApJ, 677, L9

\bibitem[\protect\citeauthoryear{{Parrish}, {Quataert} \& {Sharma}}{{Parrish}
  et~al.}{2009}]{Parrishetal09}
{Parrish} I.~J.,  {Quataert} E.,    {Sharma} P.,  2009, ApJ, 703, 96

\bibitem[\protect\citeauthoryear{{Parrish}, {Quataert} \& {Sharma}}{{Parrish}
  et~al.}{2010}]{Parrishetal10}
{Parrish} I.~J.,  {Quataert} E.,    {Sharma} P.,  2010, ApJ, 712, L194

\bibitem[\protect\citeauthoryear{{Peterson} \& {Fabian}}{{Peterson} \&
  {Fabian}}{2006}]{PetersonFabian06}
{Peterson} J.~R.,  {Fabian} A.~C.,  2006, Phys. Rep., 427, 1

\bibitem[\protect\citeauthoryear{{Pratt}, {Arnaud} \& {Aghanim}}{{Pratt}
  et~al.}{2001}]{Prattetal01}
{Pratt} G.~W.,  {Arnaud} M.,    {Aghanim} N.,  2001, in {Neumann} D.~M.,
  {Tranh Thanh Van} J.,  eds, Clusters of Galaxies and the High Redshift
  Universe Observed in X-rays: {XMM-Newton observations of galaxy clusters; the
  radial temperature profile of A2163}.
Rencontres de Moriond, Les Arcs 1800, France

\bibitem[\protect\citeauthoryear{{Rasmussen} \& {Ponman}}{{Rasmussen} \&
  {Ponman}}{2007}]{RasmussenPonman07}
{Rasmussen} J.,  {Ponman} T.~J.,  2007, MNRAS, 380, 1554

\bibitem[\protect\citeauthoryear{{Ravindranath}, {Ho}, {Peng}, {Filippenko} \&
  {Sargent}}{{Ravindranath} et~al.}{2001}]{Ravindranathetal01}
{Ravindranath} S.,  {Ho} L.~C.,  {Peng} C.~Y.,  {Filippenko} A.~V.,
  {Sargent} W.~L.~W.,  2001, AJ, 122, 653

\bibitem[\protect\citeauthoryear{{Sanders} \& {Fabian}}{{Sanders} \&
  {Fabian}}{2007}]{SandersFabian07}
{Sanders} J.~S.,  {Fabian} A.~C.,  2007, MNRAS, 381, 1381

\bibitem[\protect\citeauthoryear{{Sanderson}, {Ponman} \&
  {O'Sullivan}}{{Sanderson} et~al.}{2006}]{Sandersonetal06}
{Sanderson} A.~J.~R.,  {Ponman} T.~J.,    {O'Sullivan} E.,  2006, MNRAS, 372,
  1496

\bibitem[\protect\citeauthoryear{{Sazonov}, {Revnivtsev}, {Gilfanov},
  {Churazov} \& {Sunyaev}}{{Sazonov} et~al.}{2006}]{Sazonovetal06}
{Sazonov} S.,  {Revnivtsev} M.,  {Gilfanov} M.,  {Churazov} E.,    {Sunyaev}
  R.,  2006, A\&A, 450, 117

\bibitem[\protect\citeauthoryear{{Serra} \& {Oosterloo}}{{Serra} \&
  {Oosterloo}}{2010}]{SerraOosterloo10}
{Serra} P.,  {Oosterloo} T.~A.,  2010, MNRAS, 401, L29

\bibitem[\protect\citeauthoryear{{S\'{e}rsic}}{{S\'{e}rsic}}{1968}]{Sersic68}
{S\'{e}rsic} J.~L.,  1968, {Atlas de galaxias australes}.
Observatorio Astronomico: Cordoba, Argentina

\bibitem[\protect\citeauthoryear{{Simionescu}, {Werner}, {Finoguenov},
  {B{\"o}hringer} \& {Br{\"u}ggen}}{{Simionescu}
  et~al.}{2008}]{Simionescuetal08}
{Simionescu} A.,  {Werner} N.,  {Finoguenov} A.,  {B{\"o}hringer} H.,
  {Br{\"u}ggen} M.,  2008, A\&A, 482, 97

\bibitem[\protect\citeauthoryear{{Snowden}, {Collier} \& {Kuntz}}{{Snowden}
  et~al.}{2004}]{Snowdenetal04}
{Snowden} S.~L.,  {Collier} M.~R.,    {Kuntz} K.~D.,  2004, ApJ, 610, 1182

\bibitem[\protect\citeauthoryear{{Sun}}{{Sun}}{2009}]{Sun09}
{Sun} M.,  2009, ApJ, 704, 1586

\bibitem[\protect\citeauthoryear{{Sun}, {Jones}, {Forman}, {Vikhlinin},
  {Donahue} \& {Voit}}{{Sun} et~al.}{2007}]{Sunetal07}
{Sun} M.,  {Jones} C.,  {Forman} W.,  {Vikhlinin} A.,  {Donahue} M.,    {Voit}
  M.,  2007, ApJ, 657, 197

\bibitem[\protect\citeauthoryear{{Sun}, {Voit}, {Donahue}, {Jones}, {Forman} \&
  {Vikhlinin}}{{Sun} et~al.}{2009}]{Sunetal09}
{Sun} M.,  {Voit} G.~M.,  {Donahue} M.,  {Jones} C.,  {Forman} W.,
  {Vikhlinin} A.,  2009, ApJ, 693, 1142

\bibitem[\protect\citeauthoryear{{Tal}, {van Dokkum}, {Nelan} \&
  {Bezanson}}{{Tal} et~al.}{2009}]{Taletal09}
{Tal} T.,  {van Dokkum} P.~G.,  {Nelan} J.,    {Bezanson} R.,  2009, AJ, 138,
  1417

\bibitem[\protect\citeauthoryear{{Tonry}, {Dressler}, {Blakeslee}, {Ajhar},
  {Fletcher}, {Luppino}, {Metzger} \& {Moore}}{{Tonry}
  et~al.}{2001}]{Tonryetal01}
{Tonry} J.~L.,  {Dressler} A.,  {Blakeslee} J.~P.,  {Ajhar} E.~A.,  {Fletcher}
  A.~B.,  {Luppino} G.~A.,  {Metzger} M.~R.,    {Moore} C.~B.,  2001, ApJ, 546,
  681

\bibitem[\protect\citeauthoryear{{Trager}, {Faber}, {Worthey} \&
  {Gonz{\'a}lez}}{{Trager} et~al.}{2000}]{Trageretal00}
{Trager} S.~C.,  {Faber} S.~M.,  {Worthey} G.,    {Gonz{\'a}lez} J.~J.,  2000,
  AJ, 119, 1645

\bibitem[\protect\citeauthoryear{{Vikhlinin}, {McNamara}, {Forman}, {Jones},
  {Quintana} \& {Hornstrup}}{{Vikhlinin} et~al.}{1998}]{Vikhlininetal98}
{Vikhlinin} A.,  {McNamara} B.~R.,  {Forman} W.,  {Jones} C.,  {Quintana} H.,
   {Hornstrup} A.,  1998, ApJ, 502, 558

\bibitem[\protect\citeauthoryear{{Weisskopf}, {Brinkman}, {Canizares},
  {Garmire}, {Murray} \& {Van Speybroeck}}{{Weisskopf}
  et~al.}{2002}]{Weisskopfetal02}
{Weisskopf} M.~C.,  {Brinkman} B.,  {Canizares} C.,  {Garmire} G.,  {Murray}
  S.,    {Van Speybroeck} L.~P.,  2002, PASP, 114, 1

\bibitem[\protect\citeauthoryear{{Wise}, {McNamara}, {Nulsen}, {Houck} \&
  {David}}{{Wise} et~al.}{2007}]{Wiseetal07}
{Wise} M.~W.,  {McNamara} B.~R.,  {Nulsen} P.~E.~J.,  {Houck} J.~C.,    {David}
  L.~P.,  2007, ApJ, 659, 1153

\bibitem[\protect\citeauthoryear{{Worrall} \& {Birkinshaw}}{{Worrall} \&
  {Birkinshaw}}{1994}]{WorrallBirkinshaw94}
{Worrall} D.~M.,  {Birkinshaw} M.,  1994, ApJ, 427, 134

\bibitem[\protect\citeauthoryear{{Worrall} \& {Birkinshaw}}{{Worrall} \&
  {Birkinshaw}}{2006}]{WorrallBirkinshaw06}
{Worrall} D.~M.,  {Birkinshaw} M.,  2006, in {D.~Alloin} ed., Physics of Active
  Galactic Nuclei at all Scales Vol.~693 of Lecture Notes in Physics, Berlin
  Springer Verlag, {Multiwavelength Evidence of the Physical Processes in Radio
  Jets}.
p.~39

\bibitem[\protect\citeauthoryear{{Worrall}, {Birkinshaw}, {O'Sullivan},
  {Zezas}, {Wolter}, {Trinchieri} \& {Fabbiano}}{{Worrall}
  et~al.}{2010}]{Worralletal10}
{Worrall} D.~M.,  {Birkinshaw} M.,  {O'Sullivan} E.,  {Zezas} A.,  {Wolter} A.,
   {Trinchieri} G.,    {Fabbiano} G.,  2010, MNRAS, 408, 701

\bibitem[\protect\citeauthoryear{{Zezas}, {Birkinshaw}, {Worrall}, {Peters} \&
  {Fabbiano}}{{Zezas} et~al.}{2005}]{Zezasetal05}
{Zezas} A.,  {Birkinshaw} M.,  {Worrall} D.~M.,  {Peters} A.,    {Fabbiano} G.,
   2005, ApJ, 627, 711

\end{thebibliography}

\label{lastpage}

\end{document}